\documentclass[12pt]{article}
\usepackage{osa2}
\usepackage{overcite}
\newcommand{\citep}[1]{\cite{#1}}
\newcommand{\citet}[1]{\cite{#1}}

\newcommand{\mm}{$\,\rm{\mu m}$}
 
\newcommand{\as}{$^{\prime\prime}$}  

\newcommand{\Ohm}{\Omega}

\newcommand{\by}{$\times$}
\newcommand{\e}{\times 10^}

\newcommand{\tsys}{${\rm T_{sys}}$}

\newcommand{\aaa}{ Astron.\ Astrophys.\ }
\newcommand{\colhead}[1]{#1}
\newcommand{\nodata}{ }
\newcommand{\kkms}{$\,\mathrm{K\,km\,s^{-1}}$}
\newcommand{\kms}{$\,\rm km\,s^{-1}$}
\begin{document}
\hspace{2.0in}2002, {\it Applied Optics,} {\bf 41,} 2561-2575. 
\title{SPIFI: a direct - detection imaging spectrometer for submillimeter wavelengths}

\author{C.M. Bradford,  G.J. Stacey, M.R. Swain and T. Nikola}
\address{Department of Astronomy, Cornell University \\ Ithaca, NY 14853}
\email{bradford@submm.caltech.edu}
\author{A.D. Bolatto,  J.M. Jackson}
\address{Institute for Astrophysical Research, Boston University \\ Boston, MA 02215}
\author{M.L. Savage,  J.A. Davidson}
\address{USRA SOFIA, NASA Ames Research Center \\ Moffet Field, CA 94035}
\author{P. Ade}
\address{Department of Physics and Astronomy, University of Wales, Cardiff \\ 5 The Parade,
Cardiff, UK CF24 3YB}

\vspace{1.0in}
\hspace{2.0in}2002, {\it Applied Optics,} {\bf 41,} 2561-2575.

\begin{abstract}
The South Pole Imaging Fabry-Perot Interferometer (SPIFI) is the first instrument of its
kind -- a direct-detection imaging spectrometer for astronomy in the submillimeter band.
SPIFI's focal plane is a square array of 25 silicon bolometers cooled to 60 mK; the
spectrometer consists of two cryogenic, scanning Fabry-Perot interferometers in series
with a 60 mK bandpass filter.  The instrument operates in the short submillimeter windows
(350\mm, 450\mm) available from the ground, with spectral resolving power selectable
between 500 and 10000.  At present, SPIFI's sensitivity is within a factor 1.5-3 of the
photon background limit, comparable to the best heterodyne spectrometers.  The
instrument's large bandwidth and mapping capability provide a substantial advantages for
specific astrophysical projects, including deep extragalactic observations.  In this
article we present the motivation for and design of SPIFI and its operational
characteristics on the telescope.

\end{abstract}


\section{Introduction} 

Spectroscopy at submillimeter frequencies ($\rm \lambda \sim 100\mu m - 1 mm $) provides an
unique probe of the interstellar medium (ISM) of our galaxy and external
galaxies.  The much longer wavelength insures that dust extinction is completely
negligible relative to the near- and mid-IR lines observable from the ground.  Yet unlike
the typical mm- or cm-wave molecular transitions, most of the submillimeter lines arise
only in excited regions.  Study of the excited gas is not confused by cool molecular
material nearby or along the line of sight.  Because submillimeter lines include
rotational transitions of warm molecular gas as well as fine-structure transitions of
atomic and ionized species, the suite of lines makes an excellent tool for study of the
ionized gas / molecular gas interaction zone that results when new stars disrupt the
environment of their birth.

From the instrumentation standpoint, the submillimeter band is one of the final frontiers
of observational astronomy.  As the regime in which direct-detection and heterodyne
techniques converge, instrumentation performance has been improving rapidly during the
last two decades.  Recently, heterodyne receivers have advanced to higher frequencies,
become more sensitive, and are beginning to be assembled in small arrays
\citep{koo00,hil00}.  These receivers have been used successfully to observe submillimeter
spectral lines in a variety of galactic and a few extragalactic sources over the last
10-20 years \citet{kee85,zmu88,wal93,plu94,wil92}.  These receivers have advantages:
spectral resolution can be made arbitrarily high, and because they preserve the phase of
the radiation field, coherent receivers can be used in interferometry experiments to
obtain high angular resolution.  The preservation of phase, however, requires
amplification of the signal before detection and results in an uncertainty in photon
occupation number (intensity) of one photon per unit bandwidth per second.  When expressed
as a temperature of a radiation field which would produce the same noise, this quantum
noise is $T_{\rm RX\,QN} (SSB) = h\nu / k_B$ \citep{har99,cz96,tf85}.
In the short submillimeter bands, the quantum noise becomes comparable to the fundamental
noise due to the fluctuations in the arrival rates of photons: $T_{\rm RX,\,PN} \sim
100-200 \rm\, K$ (temperature \by\ emissivity), a fundamental limit for any detection
process (see Section~4).  Thus a coherent system is always less
sensitive than a system subject only to the photon background noise, with the difference
especially significant under low-background conditions such as air- or space-borne
observatories.

Heterodyne systems are also limited in bandwidth, especially at the higher frequencies
where the fractional bandwidth of a fixed spectrometer backend is smaller.  At present,
1.5 GHz backends are the largest used astronomically \citep{koo00}, which results in a
velocity coverage of only 560 \kms at 810 GHz.  While much larger bandwidth systems are
under development, this is currently a serious limitation for extragalactic observations
where line-widths are $\rm{\sim few \times 100\, km\:s^{-1}}$.  Often multiple spectral
setups are required to observe the full line profile \citet{gp98}.

In contrast, a direct-detection spectrometer can be background-limited and have very large
bandwidth.  Low photon energies prohibit the photon counting detectors of the mid and
far-IR \citet{sta91}, but bolometers have proven to be effective submillimeter-wave
detectors.  In continuum instruments, bolometers have approached background-limited
performance with single pixels, and recently, in large arrays such as those in the SCUBA
(Submillimeter Common-User Bolometer Array) camera in service at the JCMT\citet{hol98,wan96}.  A
background-limited spectrometer is naturally more difficult than a continuum instrument;
spectroscopy requires a large cryogenic interference pathlength, and because even 4~K
surfaces emit substantially in the submillimeter bands, careful filtration is essential.
Furthermore, the background power and corresponding photon noise on the detectors is very
small, so their detector Noise Equivalent Powers (NEPs) must be small to reach the
background limit.

We present the first direct-detection submillimeter spectrometer, the South Pole Imaging
Fabry-Perot Interferometer (SPIFI).  SPIFI uses 25 silicon bolometers in the focal plane
of an imaging system that matches each detector to the diffraction limit on the sky.  To
obtain high responsivity and low noise, the detectors are cooled to 60 mK with an
adiabatic demagnetization refrigerator (ADR).  Two cryogenic scanning metal mesh
Fabry-Perot interferometers provide spectral resolving power $R \sim \lambda /
\Delta\lambda$ between 500 and 10000 across the entire array, and careful filtration
ensures good spectral purity in the system.  SPIFI has been four years in the design,
construction and testing, and has recently undergone successful commissioning runs at the
JCMT.  The instrument is unique: it can provide a wealth of useful astronomical
observations in the coming years, and serves as an example for future instrumentation
development.  In Section~2 we describe the instrument concept and technical details.
Section~3 we discusses the operation of SPIFI on the telescope and Section~4 presents the
instrument's measured sensitivity along with a comparison with other spectrometers.

\section{Instrument Design}\label{sec:spifi-design}
\subsection{Overview}
In background-limited direct-detection spectroscopy, maximum sensitivity is obtained with
the instrumental line width matched to the instrinsic width of the astronomical line.
SPIFI is designed to observe a wide range of galactic and extragalactic targets, thus the
instrument's resolving power (R~=~ $\lambda/\Delta\lambda$) should be variable, from $c /
R$ = 30\kms\ for our Galactic Center, to 100-200\kms\ for nearby galaxies, to as low as
500\kms\ for distant galaxies.  The high-resolving-power end of this range requires a long
interference path -- R of 10,000 at 370\mm\ requires a (cryogenic) pathlength of about
$\lambda R / 2 = 1.8$ meters, so a grating spectrometer would require a huge cryogenic
volume.  A Fabry-Perot interferometer (FPI) is a multi-pass spectrometer, and a R=10,000
etalon need only be a few centimeters thick, providing a substantial benefit in a
cryogenic environment.  FPI systems are also fully tunable; the resolving power is
continuously variable by changing the cavity spacing.

Spectroscopy with an FPI lends itself naturally to a spatial imaging system, allowing
multiple detectors to be configured in a two-dimensional array on the sky.  The spatial
array is a substantial advantage for extended sources such as the Galactic Center and
nearby starburst galaxies, providing increased speed and improved registration and
calibration in the maps.  Observations of unresolved sources, such as the ULIGs and
high-redshift galaxies can also benefit from a spatial array; the source can be chopped
between two pixels on the array, improving the sensitivity by $\sqrt{2}$.  Furthermore,
off-source pixels can be used to subtract much of the ``sky noise'' due to fluctuating
telluric transmission inherent in ground-based submillimeter observations.

\subsection{Optical Design}\label{sec:optical-design}

The instrument's optical design is based on a previously-existing focal plane assembly
originally developed as an 850\mm\ photometer by collaborators J.A. Davidson and
M.L. Savage\citet{swa98}.   The 25 detectors are mounted in a square format,
separated by 6.8 mm.  The spectrometer is designed to match this detector spacing to a
$1.5\,\lambda /D$ beam on the sky at 370\mm\ and provide resolving power selectable from
500 to 10,000 over the full array.    

As with any direct detection spectrometer, a resolving power of 10,000 requires a
collimated beam through the primary dispersing element, since the maximum resolving power
$R_{\rm max}$ in a converging (or diverging) beam is $R_{\rm max} \sim {2\,\it{f}^2}$,
where {\it f} is the focal ratio.  Even with a collimated beam in the spectrometer, the
finite solid angle subtended by a diffraction-limited sky beam limits the spectral
resolution.  As the incident beam is tipped relative to the etalon normal, a Fabry-Perot
resonates at shorter wavelength according to $\lambda = \lambda_0\, \cos{\theta}$.  For
small angles, this shift is second order in the off-axis angle
($\delta \lambda \sim \theta^2$), so the spectral profile is both shifted and broadened for
off-axis fields.  The angular extent of a diffraction-limited field in the spectrometer's
collimated beam scales inversely with the collimated beam size, so the degradation of
resolution can be made negligible for a sufficiently large collimated beam inside the
instrument.  It can be shown that, to prevent resolution degradation from this effect, the
size ${D_C}$ of the collimated beam must be at least \citet{lat97,pog91}
\begin{equation}
D_C = \sqrt{2.25 \cdot \lambda^2 \cdot R \cdot n_{\rm beams}},
\end{equation}
where R is the resolving power and ${n_{\rm beams}}$ is number of diffraction-limited beams
from the center of the array to the corner.  For SPIFI, we require $R=10,000$ and with
our $5 \times $5 array, $n_{\rm beams} = 2\sqrt{2}$, thus the collimated beam inside the
instrument must be 9~cm in diameter.  

The size of the spectrometer and the optical layout presented in Figure~1 are driven by
this constraint.  The input beam is {\it f}/8.4 and reaches a focus near the dewar window.
The first mirror collimates the input beam at 9 cm, the second flat mirror sends the beam
through the High-Order Fabry-Perot Interferometer (HOFPI).  Immediately after the HOFPI,
the beam enters the $^4$He-temperature enclosure, and is directed through the Lyot stop by
flat mirror 3.  Mirror 4 re-images the field at {\it f}/12.6 for the detector array, which
is reached after flat mirror 5, the Low-Order Fabry-Perot Interferometer (LOFPI), flat
mirror 6 and the 60~mK bandpass filter.  The radiation is coupled to the 1~mm diameter
detectors with compound parabolic cones, or Winston Cones\cite{har76}, each with an
entrance aperture of 5.6 mm.  Because the cones are designed for an {\it f}/4 beam, their
response patterns are substantially oversized relative to the beam from the spectrometer, and their
only effect on the optical system is to increase the effective area of the detectors. 

All the mirrors are diamond-turned aluminum with a scratch-safe chromium-gold coating.
The two powered mirrors are off-axis paraboloids and are in a Czerny-Turner
configuration\cite{sch87} to minimize coma.  All optical elements except for the cold
aperture stop are oversized by 30\% to minimize diffraction effects from edges within the
cryogenic instrument. Since there are only 25 detector elements in the focal plane, simple
parabolic shapes are sufficient to ensure suitable optical performance.  The geometric-ray
spot sizes are less than 1~mm in the focal plane, completely negligible compared with the
5~mm diffraction spot.

\subsection{Cryogenic Design}\label{sec:cryogenic-design}
SPIFI combines two cryostats which can be operated independently, see Figure~2.  The
Fabry-Perots and spectrometer optics are housed in a large cylindrical dewar which is
split along its length to provide easy access to the optical elements when the system is
warm.  The spectrometer optics and radiation shields are cooled with liquid nitrogen and
helium cryostats constructed of welded aluminum and bolted above the aluminum optical
benches.  The 100~kg of cryogenic mass is suspended from the vacuum shell with four
1.6~mm thick $\times$ 50~mm wide $\times$ 50~mm long G-10 fiberglass struts which allow
the cryogenic assembly to remain centered with respect to the shell when cooled.  To
minimize the frequency of helium transfers, the helium cryostat occupies the bulk of the
volume above the optical benches.  Its 66 liter capacity provides a helium hold time of
five days, and with the ambient Mauna Kea atmospheric pressure, the bath operates with
$\rm T =
3.7\,K$.  The nitrogen cryostat holds 22 liters, and the vapor over the bath is pumped to 10
torr to decrease the bath temperature to 56~K.  Cooling the optics and surfaces in the
nitrogen stage lowers the background power on the detectors by 20\%, reducing its
contribution to the photon noise by 10-20\%, depending on the operating conditions.

The detector array is housed in a smaller cryostat which supports the adiabatic
demagnetization refrigerator with liquid nitrogen, $^4$He, and $^3$He systems.  The
temperature of the walls of the $^4$He-cooled enclosure is crucial, because the resulting
radiation field couples with large solid angle ($\sim$ 100 square degrees) and large
bandwidth ($\sim 1/8$) to the entrance aperture of the detector feed cones.  For this
reason, and to ensure that no stray radiation leaks into the enclosure, there are two
complete radiation shields mounted to the $^4$He bath, and the bath is always pumped to 3~
Torr, or 1.5~K so short submillimeter radiation from the surfaces is negligible.  Aside from the
$^4$He consumed in recycling the millikelvin systems, both cryostats have a hold time of more than
24 hours.  The operation of the demagnetization and $^3$He refrigerators is described in
more detail in Section~3.F.

One of the unique features of SPIFI is the joining of the two independent cryostats in a
manner which eliminates thermal infrared radiation from the beam path.  This is
accomplished with pairs of matching cylindrical extensions, ``snouts'' which butt
together, one each on the spectrometer and detector cryostats, for both the nitrogen- and
helium-temperature radiation shields.  Each pair of snouts is enclosed in a single
radiation shield which overlaps and makes a close fit.  Within these two connecting
shields, a third long cylindrical extension protrudes from the detector cryostat's inner
helium radiation shield to near the filter wheel in the spectrometer cryostat.  With this
arrangement, the stray radiation power on a detector with the instrument assembled is not
measurably different from that of the closed 1.5 K detector cryostat, to the level of
$10^{-14} \rm W$ on any given detector in our 12\% band.
 
\subsection{The Spectrometer}\label{sec:spectrometer}
\subsubsection{Triple Fabry-Perot Concept}
SPIFI obtains a resolving power R up to 10,000 with three spectral elements: two cryogenic
scanning Fabry-Perot interferometers, and one fixed bandpass filter.  The High-Order Fabry-
Perot Interferometer (HOFPI) provides the resolution of the instrument, the Low-Order
Fabry-Perot Interferometer (LOFPI), together with the fixed bandpass filter pass only the
desired order of the HOFPI,  ensuring that there is little out-of-band power transmitted by
the spectrometer, and that the in-band transmission is fairly high.  Figure~3 plots the
spectral profiles of each of these elements and the resulting system spectral profile. 

The resolving power of a Fabry-Perot etalon is the effective interference pathlength in
units of wavelength: equal to the product of the cavity order $m = 2d/\lambda$, and
finesse $F \simeq$ average number of reflections \citet{bw80}.  The partially reflecting
mirrors which form the cavity are characterized by a single pass reflection coefficient
$r$ that is typically greater than 90\%.  This determines the cavity finesse according to
$F=\pi\sqrt{r} / (1-r)$.  The transmission of the etalon is determined by the absorption
losses in cavity mirrors, characterized by a single-pass absorption coefficient for a
single mirror $a = 1 - t - r$.  For $a < 1.5\%$, the cavity transmission scales
approximately as $T_{\rm cavity} = 1 - 0.6 \cdot F \cdot a$.  Typically, our mirrors have $a
\sim 1\%$, so it is important to keep the cavity finesse modest ($F \leq 60$), in order to
maintain high transmission.

In SPIFI, the HOFPI determines the instrumental bandwidth.  It operates with finesse $F$
between 30 and 60 and order $m$ between 20 and 250.  Since the free spectral range of an
FPI is $1/m$, sorting two neighboring HOFPI orders from the desired order requires a
resolving power $R_{\rm sort} \sim m_{\rm HOFPI}$.  The broad wings of the LOFPI
Lorentzian profile require that this resolving power be somewhat greater.  As a minimum,
$R_{\rm sort} = 1.5 \times m_{\rm HOFPI}$ results in 20\% power out of band.  The LOFPI
provides this resolving power with a finesse $F_{\rm LOFPI}$ of 20-25 and order ${m_{\rm
LOFPI}}$ between 5 and 15.  The LOFPI orders must then also be sorted, for this we employ
a fixed interference filter.  These bandpass filters are constructed for SPIFI by P. Ade.
The filter shapes are more square than an FP Lorentzian profile, so that the FWHM = 12\%
filter effectively sorts LOFPI orders up to 13th with less than 15\% out of band leakage
(see Figure~3).  Both cryogenic Fabry-Perots can be tuned arbitrarily, allowing
observation of any frequency passed by the bandpass filter.  At present, the filter cannot
be changed in operation, we employ a filter centered at 366\mm, appropriate for our
primary observing program of the 370\mm\ CO and [CI] lines.  The second-generation of the
instrument under construction will allow a selection of 3-4 filters while cold.

\subsubsection{FPI Implementation}\label{sec:fpi-implementation}
Both of SPIFI's cryogenic Fabry-Perot interferometers claim heritage from the Berkeley-MPE
Far-Infrared Fabry-Perot Interferometer (FIFI)\citep{pog91}, and the Cornell Kuiper
Widefield Infrared Camera (KWIC)\citep{sta93}, far-IR spectrometers used on the Kuiper
Airborne Observatory.  The etalon mirrors are commercially available electro-formed nickel
mesh (Buckbee-Mears Corporation, St Paul, MN) which is stretched and glued onto stainless
steel rings.  For work at 370\mm, we use mesh with grid spacing of 85 and 64 \mm,
equivalent to 300 and 400 lines/inch.  The grid pattern is square, and the grid constant
and area filling factor accurately predict the single-pass reflectivity R and thus cavity
finesse F according to standard mesh theory\cite{sg83}.  Unfortunately, while the grid
constant of the mesh is well-controlled, the area filling factor cannot be specified, and
varies from batch to batch.  In practice, this prevents an arbitrary choice of finesse. 

In both scanning Fabry-Perot interferometers, each of the two mesh rings is held
magnetically to an aluminum frame.  One of these frames is mounted rigidly to the optical
bench, the other translates and so changes the cavity spacing.  The two frames are joined
in a parallelogram 'flex-vane' construction (see Figures~4 and 5).  When viewed from the side,
the moving and fixed frames form the top and bottom of a parallelogram.  At either side
are spring steel plates buttressed such that they bend only over small regions near their
connection to the frames.  This ensures that for relatively small distances ($\sim 5\%$ of
the FP size), the moving frame remains parallel to the fixed frame, and the motion does
not rely on any sliding, slipping or rolling mechanism.  The translation of this flexure
system is the scanning of the FP etalon; it is actuated with a piezoelectric transducer
(PZT) specially made for cryogenic use (Physik Instrumente (PI) Corporation, Waldbronn,
Germany).  The spacing of each of the HOFPI and LOFPI cavities is measured and controlled
with a nulling capacitance bridge circuit which set the PZT voltage\citet{lat97}.  Plate
spacing is controlled to within 0.3\mm\ with a bandwidth of a few Hz.   

Figure 4 shows the SPIFI HOFPI.  It employs the largest cryogenic PZT available, with a
cryogenic expansion range of 60\mm\ at 60 K.  At $\rm{\lambda = 370\,\mu m}$, this is
about a third of a free spectral range, or 12-20 times the spectral resolution element,
depending on the cavity finesse.  Hence the maximum scan length and the resolving power
are inversely proportional, a higher resolution setup allow smaller total bandwidth.
In addition, the HOFPI incorporates a roller-bearing translation stage actuated with a
fine adjustment screw which moves the entire inner assembly including the mirror, PZT,
flex vanes, and capacitive sensor over a 2.5 cm range.  This provides a cryogenic
adjustment range of 135 orders at 370\mm, or a change in resolving power ${\Delta R}$ from
4000 - 8000, again depending on the finesse.  Together with the roller bearing stage,
three assembly configurations make available order {\it m} up to 300 (for $\lambda =$
370\mm).  The HOFPI mesh rings have a clear aperture of 12 cm to eliminate edge effects,
and the diameter requires a thickness of 1.6 cm to carry an optically flat figure on which
the mesh can be stretched.  Once assembled and cold, the HOFPI etalon is parallelized with
three 20\mm\ PZTs, and monitored with a He-Ne laser beam reflected from both meshes.  This
technique allows parallelism to better than $3\e{-5} \rm rad$, and the etalon is generally
stable for several hours if the setup is not changed.  Whenever coarse changes are made to
the HOFPI spacing such as changing the observing bandwidth or large changes in the
frequency, the parallelism must be checked, and the voltages of the parallelizing PZTs
adjusted.

The laser retroreflection scheme also provides a means of determining the linearity of the
spectral scan.  The limitation arises in the counting of He-Ne fringes: because there are
a few $\times$ 10 optical fringes in one spectral scan, the both linearity and total length of
the spectral scan can only be determined to 2-3\%.  

The LOFPI shown in Figure~5 is somewhat smaller with a 5 cm clear aperture.  A PZT with
15\mm\ range in series with a fine adjustment screw actuate the flexure stage through a
total range of around 20 orders or 4~mm.  Because the cavity spacing is small, the meshes
themselves are used as the capacitive sensor in the LOFPI bridge.  Because it is a simpler
mechanical device, has smaller mirrors, and operates at lower resolving power, the
parallelism of the LOFPI etalon does not require cryogenic adjustment.  The mesh rings are
adjusted warm and when cold, the etalon is parallel to better than $2\e{-4}\rm\ rad$, more
than adequate for its resolving power of a few hundred.

\subsubsection{Filtration}\label{sec:filtration}

Bolometric detectors respond to all frequencies which can couple power to the absorber, so
careful attention must be paid to filtration of stray radiation, particularly in a
low-background instrument such as SPIFI.  Blocking short wavelength radiation ($\lambda$
between 5 and 100\mm) is especially critical, since at these wavelengths, the FPIs have
small but non-zero transmission and the bandpass filters invariably have leaks.  A small
leak in the filter train at short wavelength can result in substantial power on the
detector, since radiation sources warmer than 50 K emit much more power at the shorter
wavelengths than in the submillimeter.  In addition to the FPs and bandpass filter, SPIFI
employs two short wavelength blocking filters to eliminate stray
thermal infrared power.  The first, at the Lyot stop, is a flexible scatter filter on a
polyethylene substrate.  The second is a reflective mesh filter, positioned at the
entrance to the 1.5 K stage of the ADR cryostat.  Table~1 lists the elements in SPIFI's
optical and filter train, along with its transmission in our 370\mm\ band.  Tabulating
each of these values, we calculate a net transmission from the front of the instrument to
the detectors of about 0.25, a value consistent with measurements of the quantum
efficiency-transmission product.

\subsection{Bolometric Detectors}\label{sec:detectors}

SPIFI's detectors are silicon bolometers provided by S.H. Moseley and C.A. Allen at NASA
Goddard Space Flight Center, Greenbelt, MD.  They are similar to the bolometers used in
the SHARC submillimeter camera \citep{wan96} and micro-calorimeters used for x-ray
spectroscopy\cite{mcc93,mos84}.  Operational parameters of SPIFI's detectors are presented
in Table~2.  The range in values reflects both the variation in operating conditions as
spectral resolution is changed, and some variation across the array.  Each detector is
etched from a single piece of silicon and consists of a frame, four support legs, and the
detector itself, a 1~mm diameter by 12\mm\ thick disk.  The support legs provide the
thermal conductance from the detector to the frame, and two of the legs are degenerate
Boron-doped to provide an electrical connection to the thermistor.  The thermistor is a
small ($\rm{\sim 0.1\: mm^2 \times 0.5\,\mu m}$ thick) phosphorus-doped region with 50\%
Boron compensation implanted on the circular detector.  The conduction in the thermistor
is hopping mode within the valence band thus each device has a temperature-dependent
resistance given by $\rm{R = R_0 exp (\sqrt{T_0 / T})}$, with $\rm R_0 = 60-70 \Omega$, and a range
of $\rm T_0$ from 7 to 12 K.  The impedance of the dark detectors at 60 mK varies across
the array from 5 to 100 M$\Omega$, but this drops to 0.4 to 10 M$\Ohm$ as the detectors
are illuminated by the ambient background power (0.5-1.5 pW).  The thermal conductance
through the legs to the silicon frame is $G = 2-3 \times 10^{-11}$ W/K, and thus the
detectors are operating at around $\rm \Delta T = P / G =$ 75-90~mK (15-30 mK above the
60~mK bath) when exposed to the light of the spectrometer.  To increase the submillimeter
absorption, a 120$\,\mathring{\rm A}$ layer of bismuth is applied to the detector surface,
matching the surface impedance to that of free space.  For a typical detector, our
measurements indicate a quantum efficiency of 0.5, in agreement with the theoretical
prediction for a metallic film on a dielectric substrate \cite{rie94}.  Each detector is
mounted in an invar structure which forms the closed end of a 2 mm diameter integrating
cavity.  This invar cap is mounted to the base of the gold-coated copper Winston cone
which couples the radiation from the focal plane into the integrating cavity.

The bolometers are each biased in series, as shown in Figure~\ref{fig:electronics}, with a
metal film chip resistor with temperature-independent resistance of 29~M$\Omega$, mounted
directly to the back of the invar cap.  Thus the bolometers are to first order
current-biased with a typical bias voltage of 20~mV across the load resistor and bolometer
producing a detector current of a few $10^{-10}$ A.  Table~2 includes the power to voltage
responsivity of the detectors, as well as the two fundamental noise sources, Johnson noise
and thermal noise which are estimated in the standard manner\cite{rie94,mat82}.  The
magnitudes of these noise sources depend on the radiant power incident on each detector
which is a function of the resolving power, as well as on the fundamental detector
properties which vary somewhat across the array.  In addition to the Johnson and thermal
noise, the doped silicon thermistors have additional 1/f resistance noise that is
intrinsic to the conduction in the bulk thermistor, {\it i.e.}  not due to leads or
contacts \citep{han98}.  Han et al. (1998) have characterized the dependence of the noise
on the thermistor volume, temperature parameter T$_0$, and operating temperature T.  The
dependence is strong, and because it is a resistance noise, the measured voltage noise
also scales with the operating bolometer voltage.  Furthermore, the noise depends sharply
on the the thermistor doping levels which vary across SPIFI's array.  This 1/f noise
contributes between 15 and 60 $\rm nV / \sqrt{Hz}$, and is typically the dominant
non-photon noise contribution in the system.

\subsection{Millikelvin Cryogenics}\label{sec:adr}

SPIFI's bolometer array is cooled by adiabatic demagnetization of a paramagnetic salt.  The
method utilizes the fact that the thermodynamics of salt below about 1 K depend only on
the ratio of the applied magnetic field B to the temperature T, if properly
thermally isolated, a decrease in B results in decrease in T by the same ratio.  The
technique has been used for decades to cool samples in the laboratory\citet{whi79,fis98}, and
recently for astronomical X-ray spectroscopy from sounding rockets \citep{mcc96}.  We
outline our implementation here.  

The salt used in SPIFI is ${\rm FeNH_4(SO_4)\cdot 12H_2O}$, called ferric ammonium alum
(FAA).  Demagnetization of this material from a starting temperature of a pumped He bath
(1.5 - 1.8 K) can generate temperatures as low as 50 mK.  Our salt pill was constructed by
M. Dragovan at Yerkes Observatory; it is formed into a cylindrical pill of approximately
230 g and threaded with gold wires which promote good thermal conduction between the pill
and the copper detector assembly.  To minimize its heat load, the entire cold stage is
mounted with a network of kevlar threads to a $^3$He stage.  Surrounding the pill but
mounted to the $^4$He temperature structure is a superconducting solenoid magnet (American
Magnetics Corporation) which provides a central field of 4 Tesla with 20 A of current.

The cooling cycle begins by magnetizing the salt while it is thermally connected to the
$^4$He stage through a copper clamp heat switch.  
The heat of magnetization increases the temperature of the salt to around 10 K, and three
hours are required for it to return to equilibrium with the pumped He bath.  Once it has
cooled back to 1.5 K, the pill (and detector housing) are isolated from the bath by
opening the heat switch, and the magnetic field is slowly removed at a constant rate over
a period of 50 minutes.  Within 30 minutes after the demagnetization is complete, the pill
and detector housing equilibrate near 60 mK.  Access to the superconducting magnet portion
of circuit is possible through a persistence switch, a portion of the superconducting
circuit which is heated to become normal, providing a resistance in parallel with the
magnet's inductance.  A commercial automatic ramping power supply and controller are
employed for repeatable cycling of the system.
 
A thermal schematic of the refrigerator is shown in Figure~7.  Because the
demagnetization refrigerator alone is incapable of supporting the conductive loads from
1.5 K through the wiring and the mechanical mounting structure, a $^3$He system is
employed as a thermal guard.  This is a closed cycle system consisting of a condensing 
pot and charcoal pump, both with copper clamp heat switches to the $^4$He bath. 
Cycling the $^3$He system begins by isolating the pump and heating it to around 50 K to
force the gas to condense in the pot, which is thermally connected to the 1.5 K bath
through its heat switch. This process also requires three hours, both to force the $^3$He
out of the pump, and to remove the heat of vaporization of the condensing gas in the pot.
The pump is then connected to the 1.5 K bath via its heat switch, and the pot is isolated
so that the liquid $^3$He can reach its equilibrium temperature of 280 mK.  When the
instrument is at room temperature, the $^3$He resides in a built-in external storage tank
(V $\sim$ 2{\it l}), under three atmospheres of pressure.

The hold time of the 60 mK cold stage is limited by the hold time of the $^3$He guard.
The integrated heat capacity of the salt pill for a temperature change of 10 mK is 23 mJ,
which easily supports its heat loads.  Both of the dominant loads, conduction from 300 mK
through the array wiring, and radiation from the 1.5 K environment, are around 20-30 nW,
limiting the hold time to around 100 hours.  For the $^3$He stage, the dominant heat load
is conduction from 1.5 K through its fiberglass mounting structure.  This power is 75${\rm
\,\mu W}$ and vaporizes the 0.26 Joules of liquid $^3$He in about 40 hours.  In
preparation for Antarctic observations, the second generation instrument will achieve
several-day hold time by either increasing the amount of $^3$He and/or remounting the
$^3$He pot with a higher thermal-impedance structure.

\subsection{Electronics and Software}\label{sec:electronics-software}
A schematic of the bolometer amplification electronics is shown in Figure~6.  As is
customary with bolometers, the first stage for each channel is a J-FET in a
source-follower configuration.  In SPIFI the FETs are assembled on two boards with 16 FETs
each, and the boards are suspended from the ADR nitrogen stage such that the FETs heat
themselves to around 150~K.  The detectors are connected to the FETs via twisted pairs of
0.1 mm diameter constantan wire woven into ribbon cable (Oxford Instruments).  To minimize
the conductive losses to the cold stages, the cable is thermally sunk to $^3$He, $^4$He,
and N$_2$-cooled surfaces, and it is tensioned throughout to minimize microphonic pickup.  After
the J-FET buffers, the signals enter into a warm preamplifier bolted directly to the ADR
cryostat.  The signals are AC coupled with a $\rm{\tau = 1\, s}$ filter before a low-noise
G=100 amplifier, and a programmable gain amplifier.  The net gain of the cryogenic and
warm electronics is variable from 10$^2$ to 10$^5$, and the total voltage noise is 5
nV$/{\rm\sqrt{Hz}}$, referred to the detector at our chopping frequency of 8 Hz.  Because
the J-FETs are cooled, their current noise is less than $\rm{6\times 10^{-16}\,
A/\sqrt{Hz}}$, which results in a negligible ${\rm 3\, nV/\sqrt{Hz}}$ when drawn through
even a high impedance (5 M$\Omega$) bolometer.

To minimize electronic pickup, the bias voltage is provided by a battery inside the
preamplifier enclosure.  The other cryogenic and warm electronics are powered with
rechargeable lead-acid batteries, housed in a shielded box connected to the preamplifier
with a shielded cable.  After the warm preamplifier, the signals are sent via shielded,
twisted pair cables to two Keithley DAS-1800ST/HR 16-channel, 16-bit data acquisition
boards inside a PC operating Linux.  The data are processed with a custom software package
in the Matlab environment.  The software commands the Fabry-Perot positions though a
spectral scan, and at each position, co-adds the data stream from each channel with the
chop reference as a template.

\section{Operation at the JCMT}\label{sec:operation-at-jcmt}

SPIFI mounts at the right-hand Nasmyth platform, about four meters from the center of the
JCMT telescope.  The beam is delivered from the nominal {\it f}/12 focus inside the
receiver cabin to an {\it f}/8.4 focus just in front of the instrument by two lenses
mounted in the telescope elevation structure.  The lenses are diamond-machined from plate
stock blanks of high-density polyethylene (HDPE) with a refractive index of 1.52 in the
submillimeter. \citep{ash91} Both lenses are plano-convex, with diameters of 13 and 16 cm,
and mean thicknesses of 6-7 mm.  The total transmission of the lens pair is 75\%.

Measurements of the sky brightness indicate that 65\% of the power is coupled from the
telescope focus to the sky within a few arcminutes of the boresight, the remainder lost to
scattering, spillover to the ground and dome, and ohmic losses on the dish.  This defines
the forward coupling $\rm \eta_F$.  Based on our mapping of this forward beam (Figure~8),
21\% is in the main lobe, well-reproduced as a FWHM = 7.8\as\ gaussian, this is
the main beam efficiency $\rm \eta_{MB}$.  The total efficiency from the main beam in the
sky to the telescope focus is the product of these, 14\%.  The measured size
of the main lobe is somewhat larger than the nominal diffraction limit for the JCMT ($1.2
\lambda/D = 6.1^{\prime\prime}$), but both the size of the main lobe and its coupling
intensity are consistent with the other measurements at the JCMT at these
frequencies\cite{sta01a,sta01b}.  The larger beam size means that the coupling to a true point
source is even lower than the main beam efficiency, for this we define an additional
quantity $\rm \eta_A$, a measure of the fraction of the main beam which couples to a true
point source,
\begin{equation}
\rm \eta_A = \frac{\lambda^2}{A_{tel}\Omega_{MB}}. \label{eq:etaa}\end{equation} Note that
$\eta_A$ can be made unity if $\rm A_{tel}$ is decreased, it is another way of expressing
the effective aperture of the telescope.  As defined above, our measurements at the JCMT
give $\rm \eta_A = 0.48$.  The total coupling from a point source to the telescope focus is then given by
the product $\rm \eta_F\eta_{MB}\eta_A$. 

To date, SPIFI has made five observing runs at the JCMT: in April and September 1999, May and
September 2000, and May 2001.  Observations in the short submillimeter windows are
possible only in the driest conditions on Mauna Kea.  Only the first and fifth runs have
permitted useful observations at 370\mm, as the conditions during the other runs were such
that the zenith transmission was never above 10\%.  During the April 1999 run, SPIFI had
12 functional pixels, and was operating at R = 4800; we were able to map the central 2
$\times$ 3 parsecs of the Galaxy in the CO (J=7$\rightarrow$6) rotational transition with
15 pointings of the 12 pixel array, see Figure~9.  We also were able to obtain a single
array pointing of CO spectra toward the nucleus of the nearby starburst galaxy NGC 253
(see Figure~10).  These scientific results and their implications are in preparation and
will be presented in separate articles.
\subsection{Calibration}\label{sec:calibration}
Because the response of the bolometer array is a non-linear function of the bandwidth, the
instrument response must be calibrated whenever the observation setup it changed.  To
accommodate this, we employ a calibration unit permanently mounted to the front of the
instrument for both flux and spectral calibration.  A flip mirror sends the spectrometer
beam to a collimating and then a flat mirror, so that an image of the instrument pupil is
formed outside the cryostat.  At this exterior pupil is a piece of glass matched to the
beam size heated 40~K above the ambient temperature.  A rotary blade chops in front of
this heated glass, providing a flux calibration load which is identical for any position
in the focal plane.  In addition, a second rotary blade which chops immediately in front
of the cryostat entrance allows chopping the ambient blade against the sky or alignment
loads.  Absolute flux calibration of astronomical sources is always based on planets, as
they primarily couple to the main beam and not the large JCMT error beam.

For spectral calibration of SPIFI, we use CO or H$_2$S gas in a cell transited by the beam
of our exterior calibration unit.  The beam originating at the glass load transits the
50~cm gas cell twice before entering the dewar via the flip mirror, for a total path
length of about 1~m.  With this arrangement, a cell pressure of a few torr provides
sufficient column density for a high signal-to-noise ratio (SNR) absorption line spectrum
in either CO or H$_2$S.  Observation of two lines in a single scan determines the range of
the spectral scan, with a precision which depends on the SNR of the absorption features,
but is typically better than 2\%.   The scan linearity is descibed in Section~2.D.2.

Submillimeter sky transmission on Mauna Kea is monitored with radiometers operating near
200 GHz, at both the JCMT and the Caltech Submillimeter Observatory (CSO).  The measured
200 GHz opacities can be converted to a precipitable water vapor burden (PWV), the depth
of the resulting liquid layer if all the atmospheric water vapor were
condensed\cite{par01a}.  The optical depth at our observing frequency is then calculated
according most recent FTS observations and modeling of Pardo, Cernicharo and
Serabyn\cite{par01a,par01b}.  These analyses incorporate the opacity of dry air, and
provide a relationship between the PWV and the opacity at the submillimeter wavelengths.
As an example, for our primary observing frequency of 810 GHz, the opacity is given by
$\tau_{810} = 0.84 + 1.49 \times \rm PWV$.  We have verified this relationship at the JCMT
by measuring the emission of the sky as a function of zenith angle and fitting the result
to a uniform-temperature atmosphere.  With this relationship, sky dips are not needed as
long as the 200 GHz radiometers are in operation.

\section{Sensitivity}\label{sec:sensitivity}
The natural measure of sensitivity in a direct-detection system is the noise equivalent
power NEP [${\rm Watt/\sqrt{Hz}}$] or noise equivalent flux density NEFD [${\rm
Jy/\sqrt{Hz}}$] ($1\, {\rm Jy = 10^{-30}\, W\, cm^{-2}\, Hz^{-1}}$).  These quantities are
defined in terms of the SNR, easily measured with a calibration load or astronomical
source.  The relationship is given by 
\begin{equation} 
\rm SNR = \frac{P_{obs} \sqrt{2 t_{int}}}{NEP},\label{eq:nep}
\end{equation}
where $P_{\rm obs}$ is the signal power, $t_{\rm int}$ is the integration time, and the
$\sqrt{2}$ is according to the convention for expressing the NEP with units of $\rm W /
\sqrt{Hz}$.  The NEP is a general quantity which can refer to power on the detector, at
the front of the instrument, or in the main beam on the sky.  The NEFD is generally
referred to a point source above atmosphere.  There are few points sources with sufficient
flux for an 800 GHz spectrometer, but planets tend to fill the main beam, allowing their
coupling to be measured.   Taking $\rm NEP_{MB}$ to refer to the main beam, we write
\begin{equation}
\rm NEFD=\frac{NEP_{MB}}{\eta_{A} A_{tel} \Delta\nu}.
\end{equation}
where $\rm \eta_A$, given in Equation~(\ref{eq:etaa}), distinguishes between a source
filling the main beam and a true point source, and the bandwidth $\rm \Delta \nu$ is the
instrumental bandwidth.

It is customary, however, to describe heterodyne sensitivities with temperature units.  To
compare SPIFI with a heterodyne system, we convert the NEFD into a system temperature
\tsys.  

The flux density sensitivity,
$\rm S(rms)$ is related to the RMS antenna temperature $\rm T_A^* (rms)$ through
\begin{equation}
\rm S(rms) = \frac{2\,k_B\,T_A^*\,(rms)}{\lambda^2}\, \Omega_{beam} \, \frac{1}{\eta_{MB}}.\label{eq:srms}
\end{equation}
Here $\rm T_A^* (rms)$ is given by the radiometer equation 
\begin{equation}
\rm T_A^*\,(rms) = \frac{2\,T_{sys} \,\kappa}{\sqrt{\Delta \nu \cdot t_{int}}} \label{eq:rad},
\end{equation}  where the 2 accounts for chopping, $\kappa$ is the backend degradation factor,
equal to 1.15 for correlator spectrometers.\cite{mat99}
Since S(rms) and the NEFD are essentially the same, $\rm S(rms)= NEFD/\sqrt{2\,
t_{int}}$ from Eq.~(\ref{eq:nep}), we have
\begin{equation}
\rm{T_{sys}} = \frac{{\rm NEP_{MB}}\,\eta_{MB}}{4\sqrt{2}\, k_B\, \kappa \sqrt{\Delta\nu}}. \label{eq:tsys}
\end{equation}  
Unlike the NEP and NEFD, the system and RMS antenna temperatures ($\rm T_{sys}$, $\rm
T_{A}^{*}$) refer to the total coupling to the sky, {\it i.e.} the total efficiency which
is not coupled to the ground, the dome, supports, etc.  Thus the main beam efficiency, is
included in Equation~(\ref{eq:tsys}),

The values of the NEP, NEFD, and \tsys\ for SPIFI on the JCMT are presented in
Table~\ref{tab:sens}, tabulated for several resolving powers.  Sensitivities at R=4800 and
R=1500 have been measured at the telescope, though under poor atmospheric conditions
of 5-10\% zenith transmission.  Values presented are scaled to the conditions of a
good night at zenith, 40\% transmission, by simply scaling the observed SNR with sky
transmission.  The sensitivities at other resolving powers are estimated by combining the
photon noise with the excess voltage noise and optical responsivity:
\begin{equation}
\rm NEP^2 = NEP^2_{photon} + \left(\frac{NEV_{excess}}{S_{opt}}\right)^2,\label{eq:4}
\end{equation}
where the $\rm NEP_{photon}$ is calculated theoretically (see below).  The
excess noise and optical responsivity are functions of power on the detector and bias
current, and are estimated from our load curves and noise measurements.  Because the 1/f
noise and the optical responsivity both increase with resolving power, the second term on
the RHS of Eq.~(\ref{eq:4}) varies slowly, less than a factor of two over the range of
resolving powers listed.

\section{Photon Background Noise}\label{sec:phot-backgr-noise}
To evaluate the instrument's performance, it is instructive to compare the measured
sensitivities with the fundamental limit imposed by the fluctuations in the photon arrival
rates, the background limit.  The uncertainty in power $\sigma^2$ due to the photon
background after an integration time $t_{\rm int}$ can be expressed as\citet{lam86,zmu00,ben98}
\begin{equation}
\rm \sigma^2 = \frac{1}{t_{int}} \sum_i \int d\nu (h\nu)^2\left (n_i\,(T,\nu)\,\epsilon_i
\,\tau_i\,\eta\right)\, \left(n_i\,(T,\nu)\,\epsilon_i\,\tau_i\,\eta+1\right).
\label{eq:sigma} \end{equation}
The sum is over all spatial and polarization photon modes $i$ which couple to the
detector, each with total coupling given by the product of emissivity $\epsilon_i$,
transmission $\tau_i$ and detector quantum efficiency $\eta$.  The $n_i$ are the usual
photon mode occupation numbers, $n_i = \left(\exp{(h\nu/kT_i)} - 1\right)^{-1}$.  We consider only the
noise generated by the ambient radiation outside the instrument, which couples only in a
small bandwidth and with a well-defined throughput:
\begin{equation}
\rm \sigma^2 = \frac{(h \nu_0)^2}{t_{int}} \,N_{m}\,\Delta\nu \left( n_{a,\nu_0}\, \epsilon_{amb}\,
\tau \,\eta \right)
\left( n_{a,\nu_0}\, \epsilon_{amb}\, \tau\, \eta + 1 \right)
\end{equation}
where the frequency interval has been approximated because the fractional bandwidth is
small, equal to $\pi/2$ times the resolution bandwidth due to the Lorentzian line shape of
the Fabry-Perot.  Because SPIFI's optical train couples only a single spatial mode from ambient
temperature to each detector, the number of modes $\rm N_m$ is 2, including both polarizations.

The emissivity of the ambient radiation $\rm
\epsilon_{amb}$ includes contributions from the sky, telescope, and re-imaging lenses:
\begin{equation}
\epsilon_{\rm amb} = \left(1-t_{lens}\right) +
t_{lens}\left[\left(1-\eta_{F}\right)+\eta_F\left (1-t_{sky}\right)\right],
\end{equation}
where $t_{\rm lens} = 0.75$ is the total transmission of two polyethylene re-imaging
lenses, $\rm \eta_F$ is the total forward coupling of the telescope, 0.65 at 800 GHz
on the JCMT, and $t_{\rm sky}$ is the transmission of the sky.  On a good night at Mauna Kea, the
zenith sky transmission is 40\%, resulting in a total emissivity of the ambient radiation of
$\rm \epsilon_{amb} = 0.81$.   Other factors in the expression are the mode occupation number
$n_{\rm amb,\nu_0}$ for which we take $\rm T_{amb}=280\, K$, the instrument transmission
$\tau = 0.25$ and detector quantum efficiency $\eta = 0.5$.  Combined together, the product
$\rm \gamma = n_{a,\nu_0}\,\epsilon_{amb}\,\tau\,\eta$ measures the degree to which photons are
correlated at the detector.  For $\gamma < 1$, the photons are uncorrelated, $\sigma
\sim \sqrt{\gamma}$, and the noise obeys Poisson statistics as with optical and infrared
instruments.  For $\rm \gamma > 1$, $\rm \sigma \sim \gamma \sim P_{tot}$; the photons are
completely correlated and the uncertainty scales as the total incident power as with a 
traditional radio receiver.  SPIFI operates near the threshold between these two
regimes, with $\gamma$ between 0.7 and 0.8.  

To convert the uncertainty $\sigma$ to a measured SNR, we note that when chopping between
a source and the background, the effective on-source integration time is half of the total
integration time, and that the noise is $\sqrt{2}$ larger since the background and source
both contribute independent noise.  Thus we have a measured signal to noise of 
\begin{equation}
\rm SNR = \frac{P_{amb}}{\sqrt{2} \sqrt{2} \cdot \sigma},
\end{equation}
\noindent and we define the photon background-limited NEP according to Eq.~(\ref{eq:nep}),
\begin{equation}
\rm NEP_{photon} = \frac{P_{amb}\,\sqrt{2\,t_{int}}}{SNR} = 2 \sqrt{2} \, \sigma (1 sec)
\,\,\, \left[ \frac{W}{\sqrt{Hz}} \right].
\end{equation}
These values are referred to the detector, and include the detector quantum efficiency. To
refer the NEP to the sky, this value is simply divided by the total coupling between the
detector and the sky, which under good atmospheric conditions, is 0.5 (quantum efficiency)
$\times$ 0.25 (cryogenic instrument) $\times$ 0.75 (re-imaging lenses) $\times$ 0.65
(telescope forward coupling) $\times$ 0.4 (atmospheric transmission)
= 0.024.  Calculating this photon-limited NEP for the perfect system, referring it to the
sky, and converting it into a system temperature gives \tsys$\,_{\rm PBGL}$ = 1450 K, see
Table~\ref{tab:sens}.  

For any real system, the true photon noise contribution is greater than what would be
calculated according to the above method.  Because the photons are somewhat correlated in
SPIFI, an increase in the ambient emissivity from 0.81 to 1 (23\%), as was the case when
our measurements were made, results in an increase in the photon NEP by 17\%, giving a
\tsys\ of 1700 K.  Furthermore, out-of-band leakage power from ambient temperature
outside, and in-band radiation from inside the instrument also couple to the detectors and
contribute to the noise.  In SPIFI, the portion of the spectrometer at 60 K contributes
power with very low emissivity, approximately 10\%, but with a larger bandwidth since it
is not filtered by the HOFPI.  Depending on operating parameters related to the resolving
power, namely the HOFPI and LOFPI orders and finesses, this 60 K radiation contributes
between 15 and 50\% of the total detector power, and an additional 8--20\% of the
background noise.  When this portion is included, the true photon-noise \tsys, including
scanning for SPIFI is around 1830--2040 K, a large portion of what is observed,
particularly at the lower resolving powers.  As the resolving power increases, both the
background power and signal power on the detector decrease and the intrinsic detector
noise becomes more important.

As shown in Table~\ref{tab:sens}, the system is within a factor
of 1.5 - 3 of $\rm T_{sys} {\rm(PBGL)}$, the background limit for an perfect spectrometer with
25\% transmission and 50\% detector quantum efficiency.  To compare directly with a
heterodyne spectrometer, we convert the \tsys\ into a $\rm T_{RX}$ using the heterodyne
conventions \citep{mat99}:
\begin{equation}
\rm T_{sys}=2\,\, \frac{T_{RX} (DSB) + \eta_{F}\,(1-\eta_{sky})\,T_{sky} +
(1-\eta_{F})T_{\rm }}{\eta_{sky} \eta_{F}}, \label{eq:5}
\end{equation}
\noindent where $\rm \eta_{F}$ and $\rm T_{tel}$ are the total forward coupling and
physical temperature of the telescope, taken to include the losses in the polyethylene
lenses, $\rm \eta_{sky}$ and $\rm T_{sky}$ are the transmission and physical temperature
of the sky.  Our measured system temperatures over the range of resolving powers, imply
equivalent $\rm T_{RX}$ (DSB) in a range from 40 to 130 K.  The best reported 800 GHz
heterodyne receiver temperature to date is 205 K \citep{koo00b}, somewhat higher.

Equation~(\ref{eq:5}) illustrates the difference between heterodyne and direct-detection noise
conventions -- the expression of noise in a direct-detection system (eg NEP, NEFD)
includes contributions from all noise sources, including the photon background, while the
receiver temperature $\rm T_{RX}$ represents only the additional noise generated in the
receiver.  In the model perfect background limited system outlined above (with reimaging
lenses) for which the photon NEP results in a system temperature of 1450 K, the sky and
telescope terms $\rm \eta_{F} (1-\eta_{sky})T_{sky} + (1-\eta_{F})T_{tel}$ in
Eq.~(\ref{eq:5}) would sum to $0.82 \times 280\,\rm K$ (see Eq.~11).  The equivalent DSB
receiver temperature is a nonsensical -84 K.  The negative receiver temperature is due to
the fact that the heterodyne conventions account for background and receiver noise
coupling into both heterodyne sidebands, not appropriate for a direct-detection
spectrometer.  Without the factor of $2$ in Eq.~(\ref{eq:5}), the receiver temperature
based on the photon-limited NEP is +57 K, greater than zero because the Fabry-Perot
couples a larger noise bandwidth than the resolution bandwidth to which the
sensitivity is referred.

Because SPIFI is a Fabry-Perot system, we must scan a few resolution elements to
produce a spectrum, this raises the effective system temperature by $\rm \sqrt{n}$.  For a
five-element scan, the factor of just over 2 in the system temperature results in a
substantial change in the effective receiver temp, giving $\rm T_{RX}$ (DSB) ranging from
270--570 K.  This points out the disadvantage of a Fabry-Perot system, it couples only a
single resolution element of bandwidth at at time.  A grating spectrometer, by contrast,
can couple many resolution elements simultaneously.  If background-limited, a grating would
be the most sensitive spectrometer possible for point sources.  For observations of
extended sources, however, a Fabry-Perot spectrometer such as SPIFI provides instantaneous
two-dimensional mapping capability, which offsets the disadvantage of the single
instantaneous spectral element.

\section*{Acknowledgments}
We are indebted to S.H. Moseley and C.A. Allen at NASA Goddard Space Flight Center for
providing SPIFI's bolometers.  We thank several undergraduate students at Cornell
University for their diligent effort.  Chuck Henderson, Wayne Holland, and the JCMT staff
and operators deserve credit for their help interfacing SPIFI with the telescope, and we
acknowledge Antony Stark and Peter Hargrave for their support during our first observing
run.  Finally, we thank an anonymous referee for several helpful suggestions.  This work
was made possible with NASA grants NAGW-4503 and NAGW-3925 and NSF grants OPP-8920223 and
OPP-0085812.

\newpage

\section*{List of Figures}

\begin{figure}[h]
\caption{SPIFI Optical Layout.  The beam is near its {\it f}/8.4 focus as it enters the
cryostat through the polyethylene window.  M1 is an off-axis paraboloid and collimates the
beam at 9cm, M2 is a flat mirror that directs the collimated beam through the High-Order
Fabry Perot Interferometer (HOFPI) and into the 3.7 K enclosure.  Flat mirror M3 directs
the beam through the cold aperture stop (Lyot stop) and scatter filter to the camera
mirror M4, another off-axis paraboloid.  The field is imaged at {\it f}/12.6 on the
detector array via flat mirrors M5 and M6 after the beam transits the low order Fabry
Perot Interferometer (LOFPI), and mesh filters at 1.5 and 0.06 K.  For clarity, M6 and the
detector array are shown rotated 90$^\circ$ about the beam axis as it is incident on M6
(see also Figure~2). \label{fig:optic}}
\end{figure}

\begin{figure}[h]
\caption{SPIFI Cryostat.  Two independent cryostats joined together share a common vacuum
and radiation shields.  In the spectrometer cryostat, the liquid nitrogen and
helium-cooled optical benches lie below the cryogen vessels.  A lengthwise split in this
large vacuum vessel provides easy access to the optical components when it is inverted.
The smaller detector cryostat supports the millikelvin refrigerator (see Figure~7), with
liquid nitrogen and pumped liquid helium baths.  The system of matching cylindrical snouts
and close-fitting connecting shields excludes ambient radiation to the level of 1\% of the
power in the spectrometer beam ($\sim 10^{-14}$ W per detector in the 12\% band.
\label{fig:dewar}}\end{figure}

\begin{figure}[h]
\caption{SPIFI Spectrometer -- Example Configuration for R=6000.  The spectral profiles of
the three components in the spectrometer are plotted for an example setup (each offset and
plotted at 1/4 vertical scale).  The net profile is the product, plotted with the thick
curve at the bottom (full vertical scale).  The peak transmission of 0.41 does not include
losses in the window, mirrors, or the thermal IR blocking filters. \label{fig:spec}}
\end{figure}

\begin{figure}[h]
\caption{SPIFI High-Order Fabry Perot Interferometer (HOFPI) with the outer side walls
removed.  The FPI mirrors are nickel mesh stretched and glued onto stainless-steel rings (shown in
section) held magnetically to the fixed and moving frames.  To adjust the cavity order $m$
(here based on $\rm \lambda = 370\,\mu m$), the entire inner assembly is translated on the
roller-bearing stage via the adjustment screw.  For spectral scanning, the PZT pushes the
inner frame to the left against the bottom of the flex vane parallelogram.  The top of the
flex-vane parallelogram is part of the inner frame; the bottom remains fixed to the outer frame
with the adjustment screw.\label{fig:hofpi}}
\end{figure}

\begin{figure}[h]
\caption{SPIFI Low-Order Fabry Perot Interferometer (LOFPI) with the side walls removed. \label{fig:lofpi}}
\end{figure}

\begin{figure}[h]
\caption{SPIFI Electronics Block Diagram.  Each of the bolometers is biased in
series with a 29 M$\Omega$ load resistor, and buffered with a cooled J-FET.  To minimize
pickup, all signal wires are twisted pairs or coaxial, the FETs and amplifiers are powered
with batteries, and the preamplifier box is bolted tight to the cryostat.  Also shown are
the essentials of the capacitance bridge circuit and its connection to the Fabry-Perot
interferometers.  Dashed lines indicated shielded cables.  \label{fig:electronics}}
\end{figure}

\begin{figure}[h]
\caption{Adiabatic Demagnetization Refrigerator Schematic.  A network of Kevlar threads
support the 60 mK stage and its 300 mK thermal guard.  Mechanical heat switches
connecting the salt pill and $^3$He pot to the 1.5 K bath are closed when recycling the
systems.  The $^3$He pump heat switch is opened and the pump is heated for the recycling,
forcing $^3$He to condense in the pot.  In operation, the heat switches are reversed.  The
hold time of the system is around 40 hours, limited by conduction through the fiberglass
mounting of the $^3$He pot.\label{fig:adr}}
\end{figure}

\begin{figure}[h]
\caption{Integrated intensity map of the CO $J=7\rightarrow6$ (370\mm) emission from the
Galactic Center Circumnuclear Disk (CND).  Offsets are in arc-seconds relative to the
dynamical center of the galaxy SGR A$^*$ ($\rm{RA_{1950} = 17^h\,42^m\,29^s.3}$,
$\rm{Dec_{1950} = -28^\circ\, 59^{\prime}\,19^{\prime}}$), and the beam size is 11\as\
FWHM.  Emission traces warm, dense molecular gas and is observed throughout the the map;
contours are linear in $\rm{T_{MB}\Delta v}$, with a 250\kkms\ interval.  The peak in the
southwest CND is 5400\kkms, the minimum at the edges of the map is 750\kkms.  Velocity
resolution in the spectra is 50$\,{\rm km\,s^{-1}}$.  Overall rotational motion is evident in the shifting
of the velocity from north to south.
\label{fig:gc}}\end{figure}

\begin{figure}[h]
\caption{CO (J=7$\rightarrow$6) in the starburst galaxy NGC 253.  These data are a single
footprint of the array from our first run at the JCMT in April 1999, and were obtained in
15 minutes of on-source integration time with atmospheric transmission to the source
between 5-8\%.  The data indicate widespread excited molecular gas in the nuclear starburst of NGC
253. \label{fig:n253}}
\end{figure}

\clearpage

\newpage
\centerline{\scalebox{1.}{\includegraphics{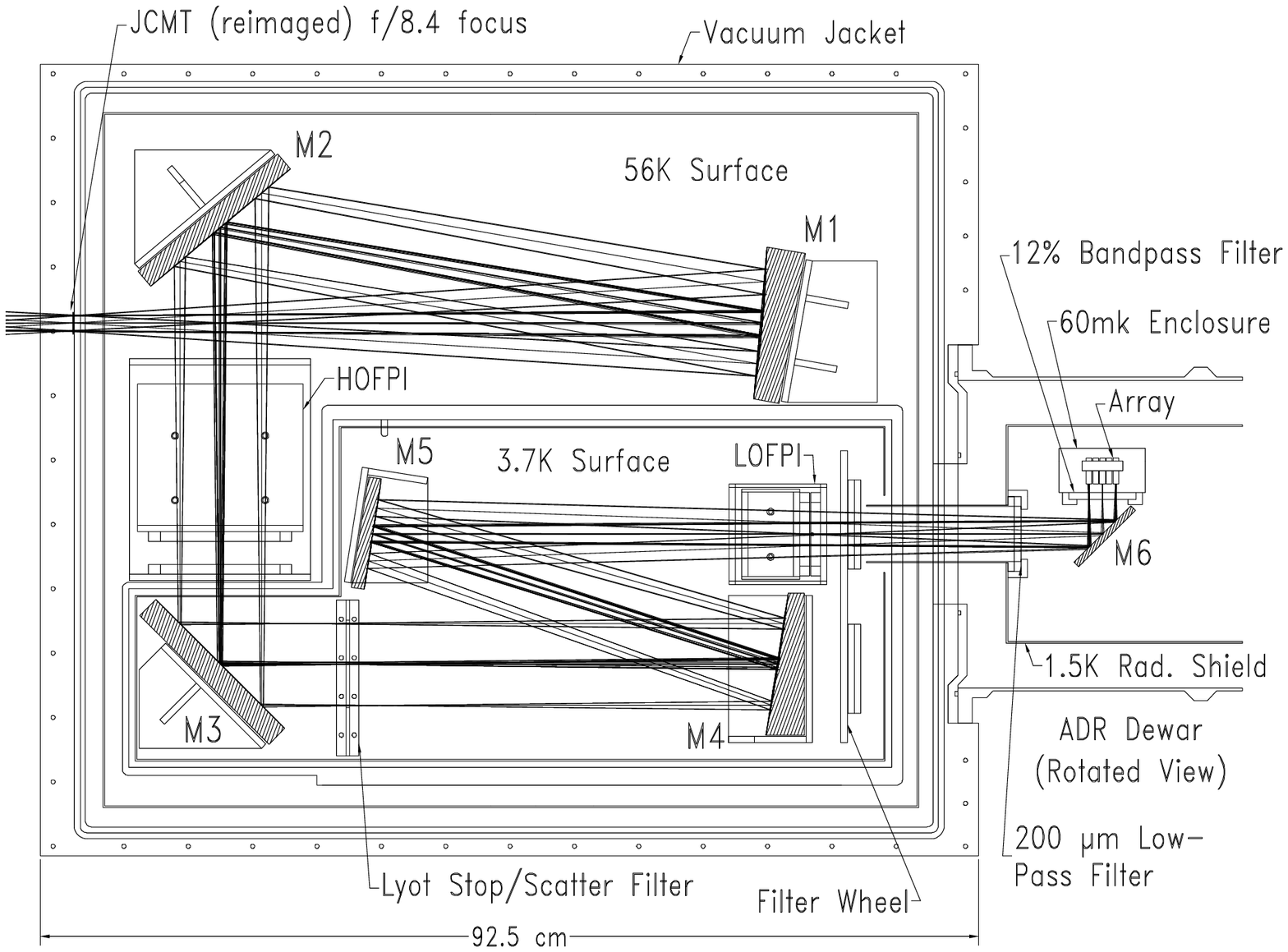}}}
\vskip2in
Figure 1, Bradford et al.  (probably will be only a single column wide)

\newpage
\centerline{\scalebox{1.}{\includegraphics{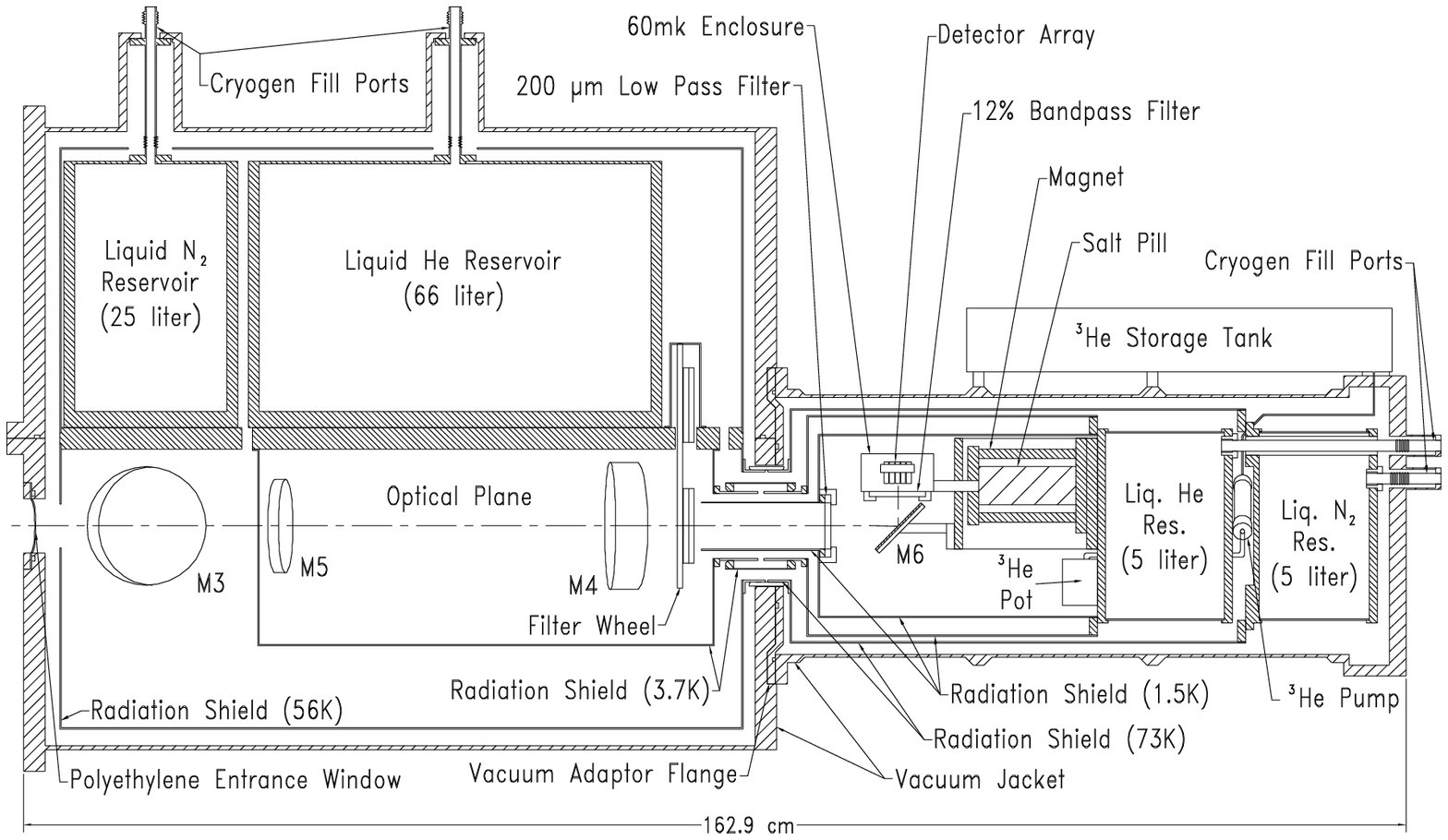}}}
\vskip2in
Figure 2, Bradford et al.   (would like full page wide)

\newpage
\centerline{\scalebox{1.}{\includegraphics{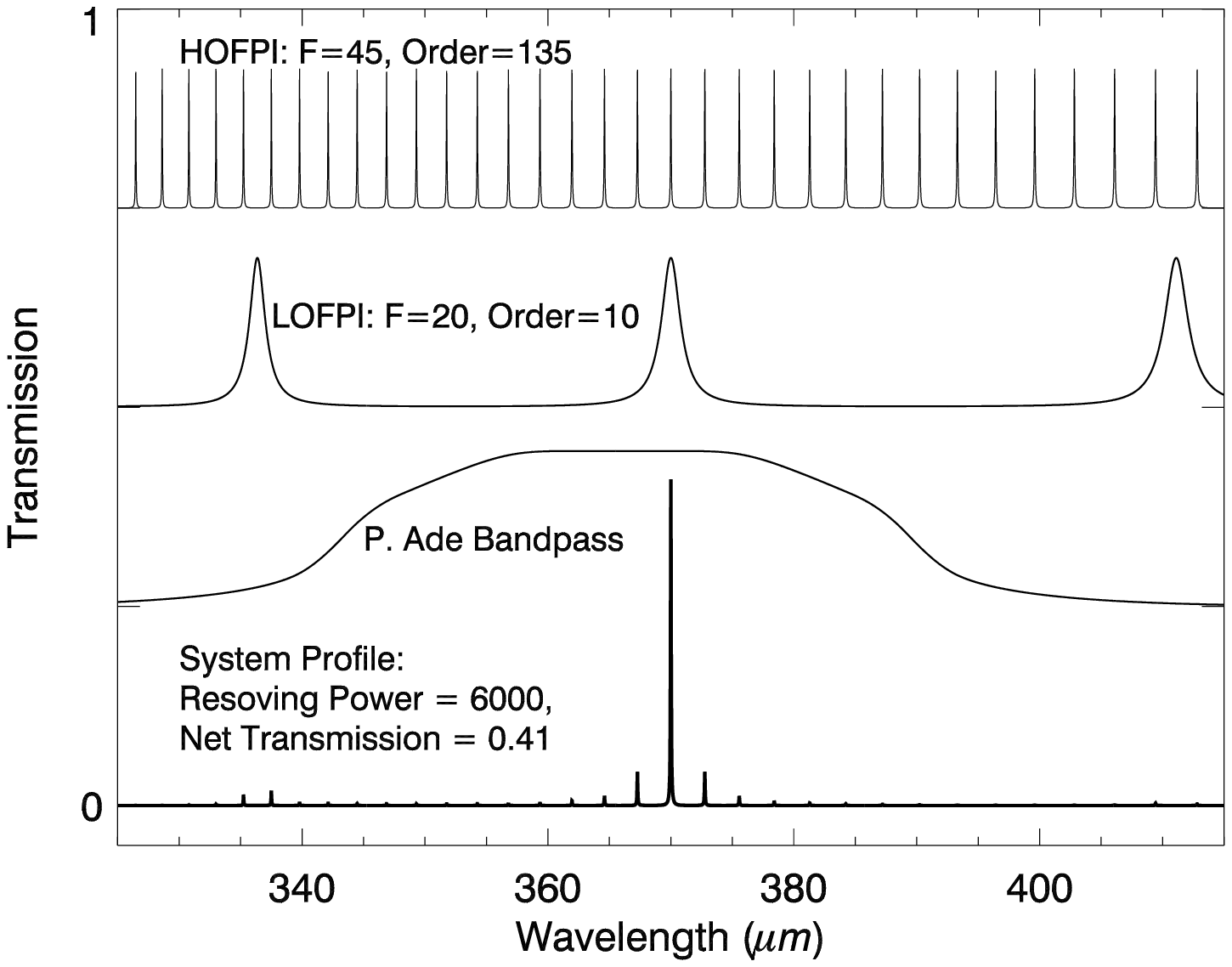}}}
\vskip2in
Figure 3, Bradford et al.   (single column)

\newpage
\centerline{\scalebox{0.8}{\includegraphics{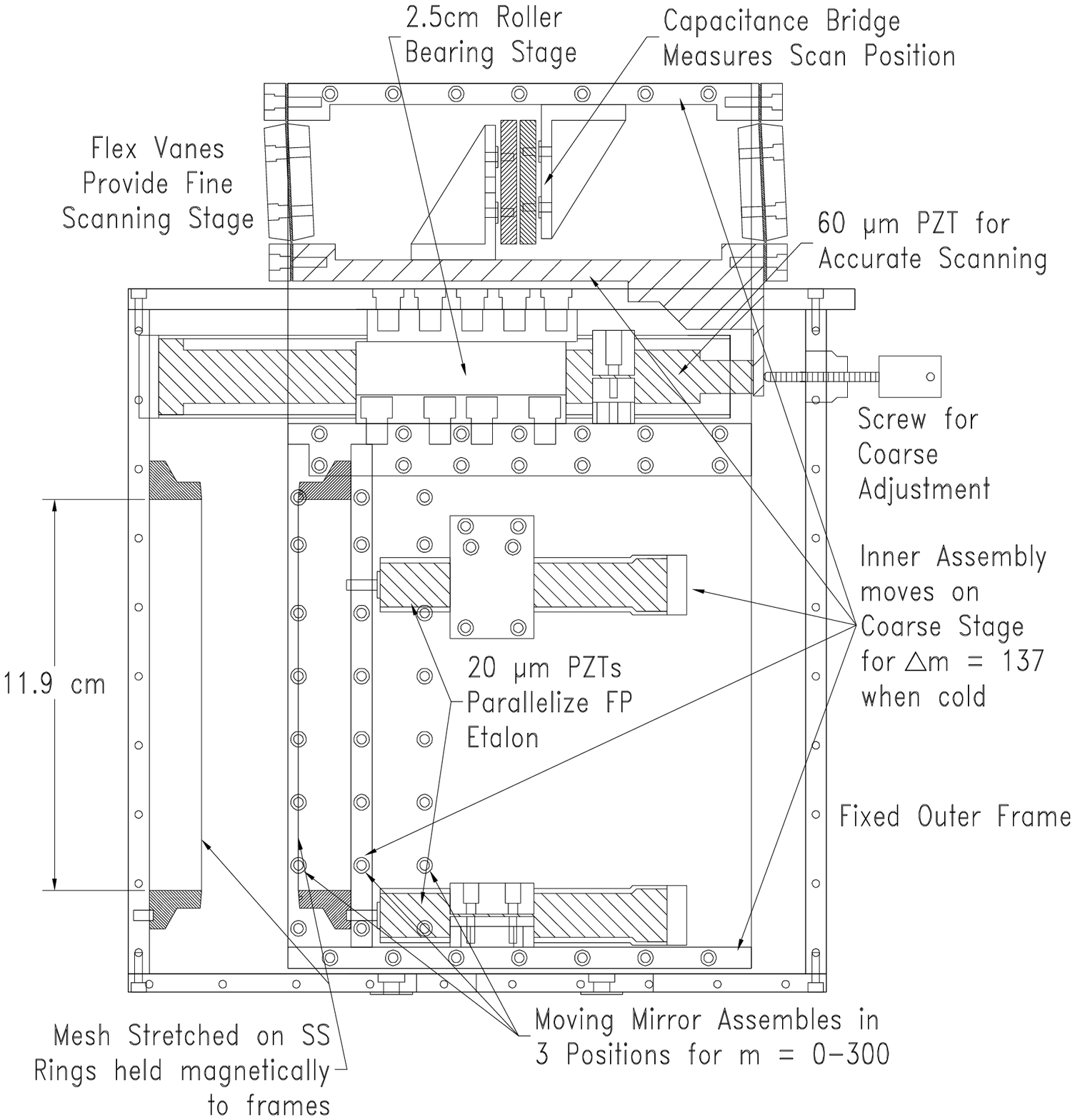}}}
\vspace{1.5in}
Figure 4, Bradford et al.   (single column)

\newpage
\centerline{\scalebox{0.6}{\includegraphics{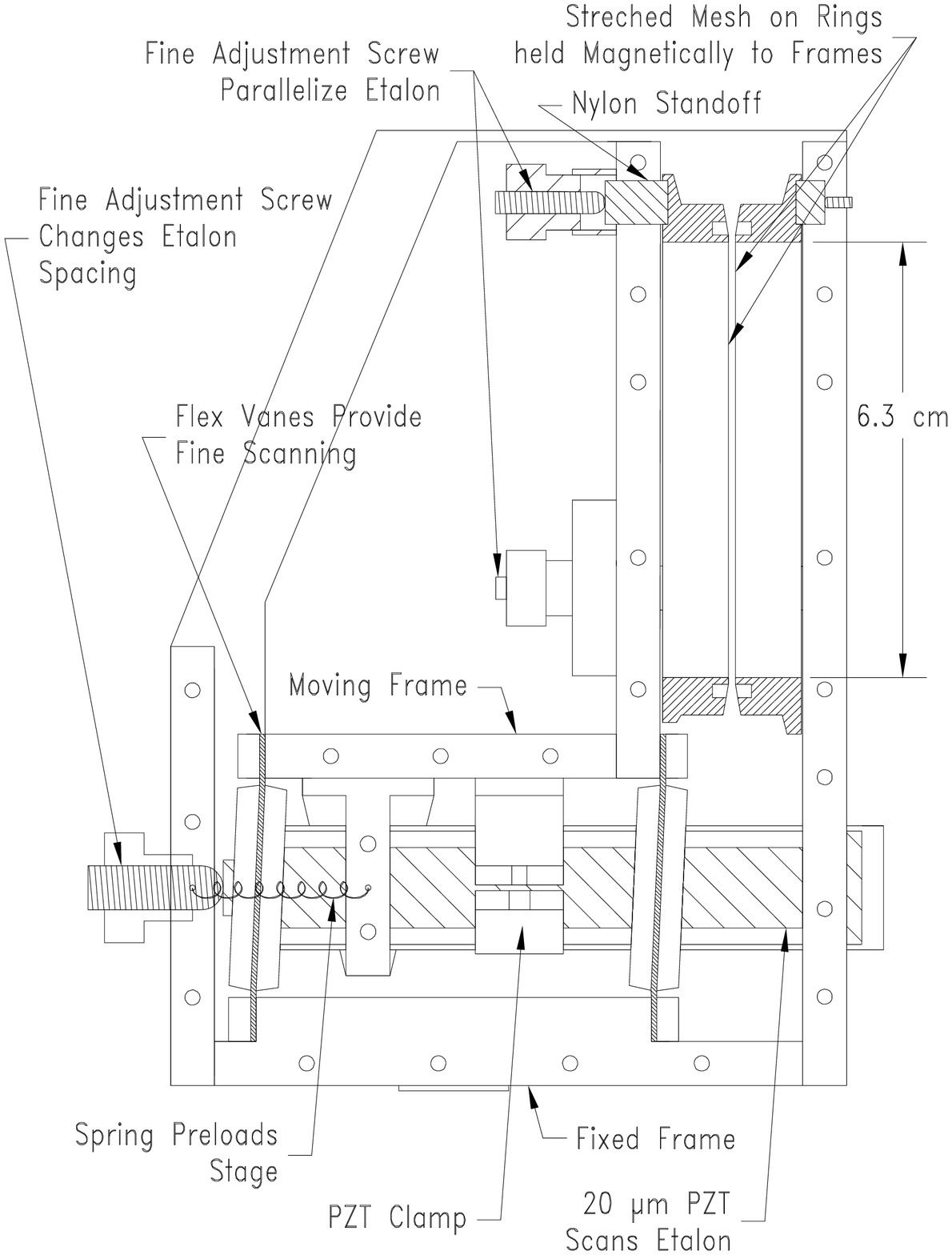}}}
\vspace{1.0in}
Figure 5, Bradford et al. (single column)

\newpage
\centerline{\scalebox{0.8}{\includegraphics{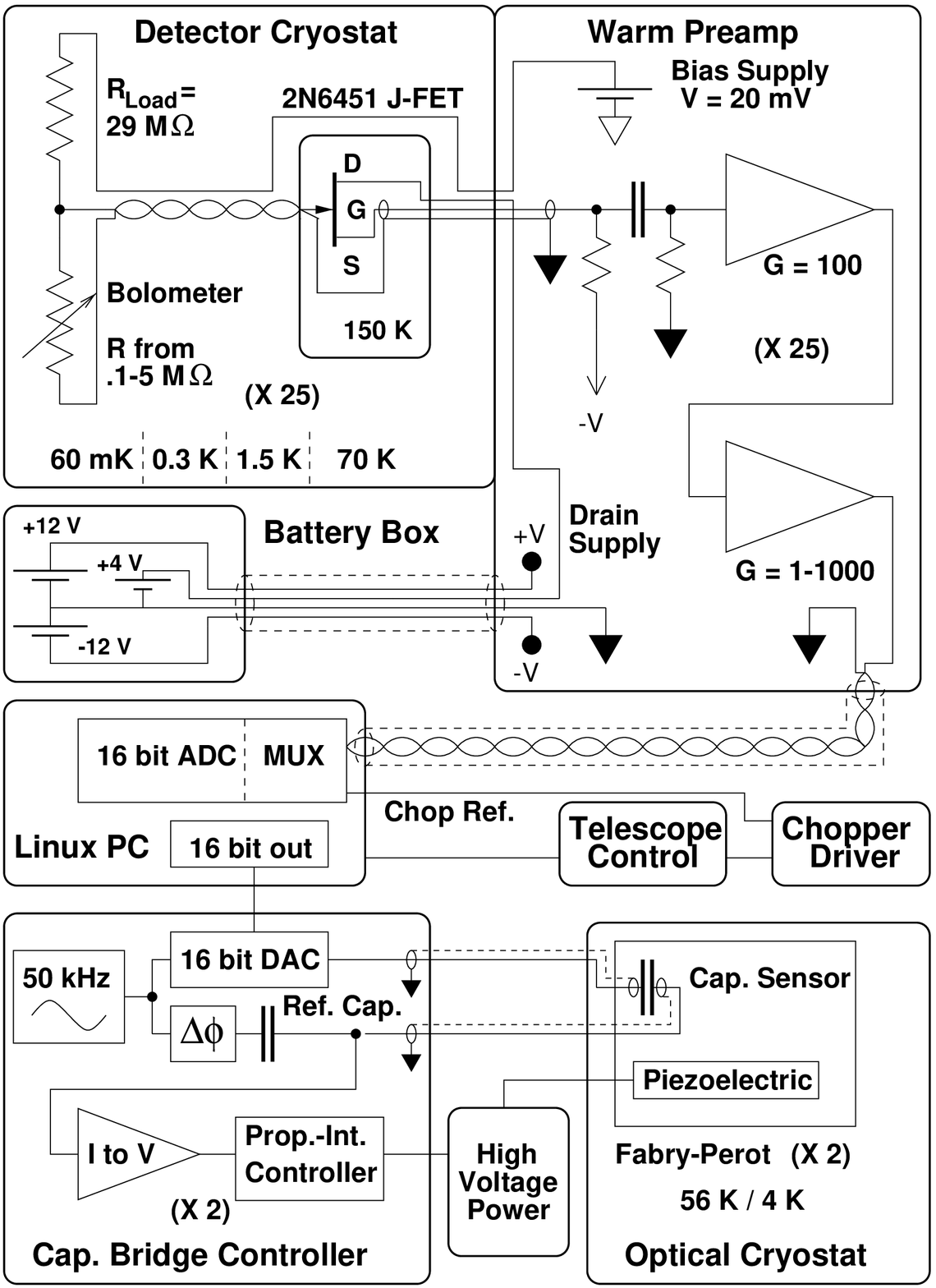}}}
\vspace{1.0in}
Figure 6, Bradford et al.   (single column)

\newpage
\centerline{\scalebox{1.}{\includegraphics{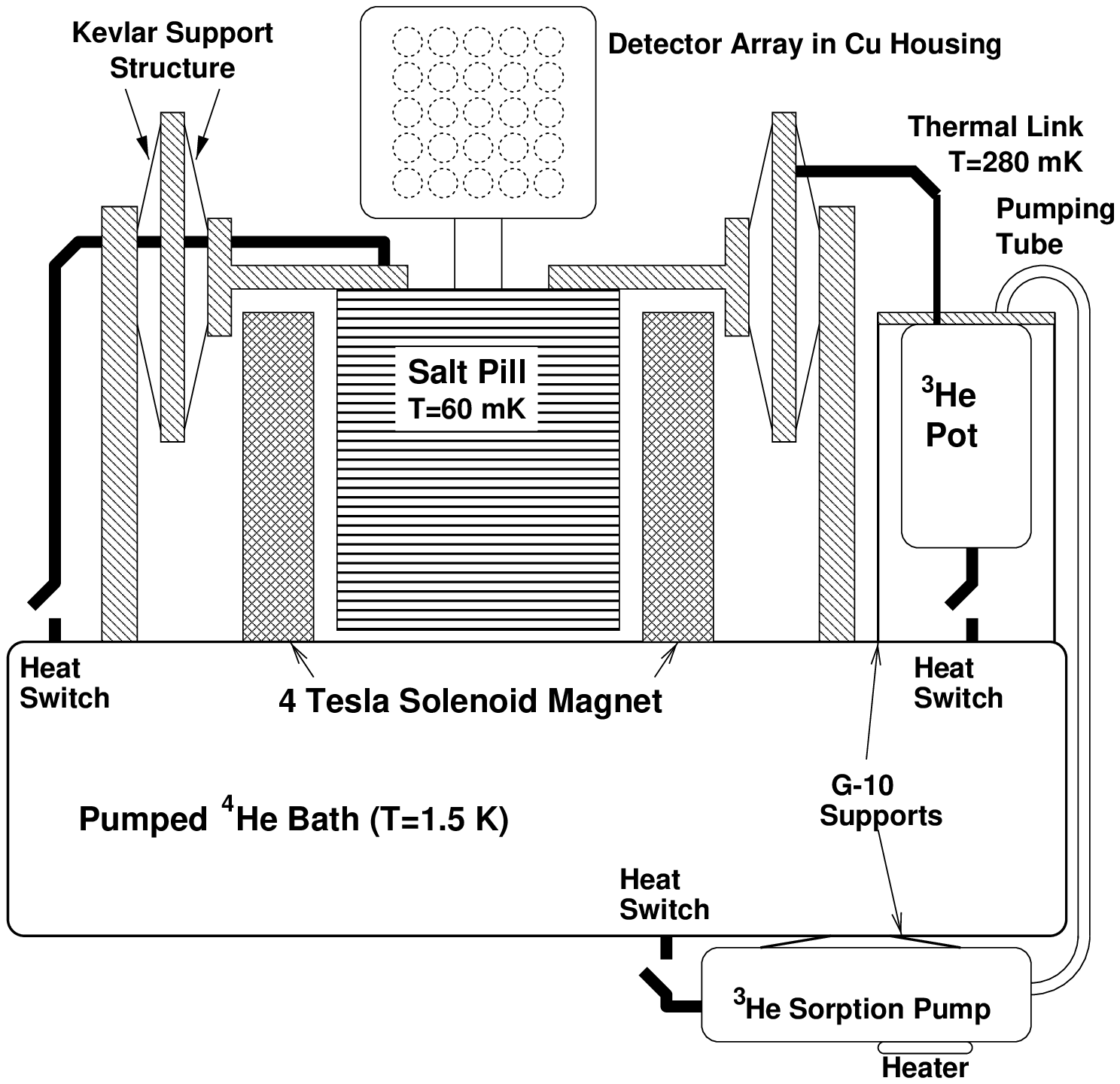}}}
\vskip2in
Figure 7, Bradford et al.   (single column)


\newpage
\centerline{\scalebox{1.}{\includegraphics{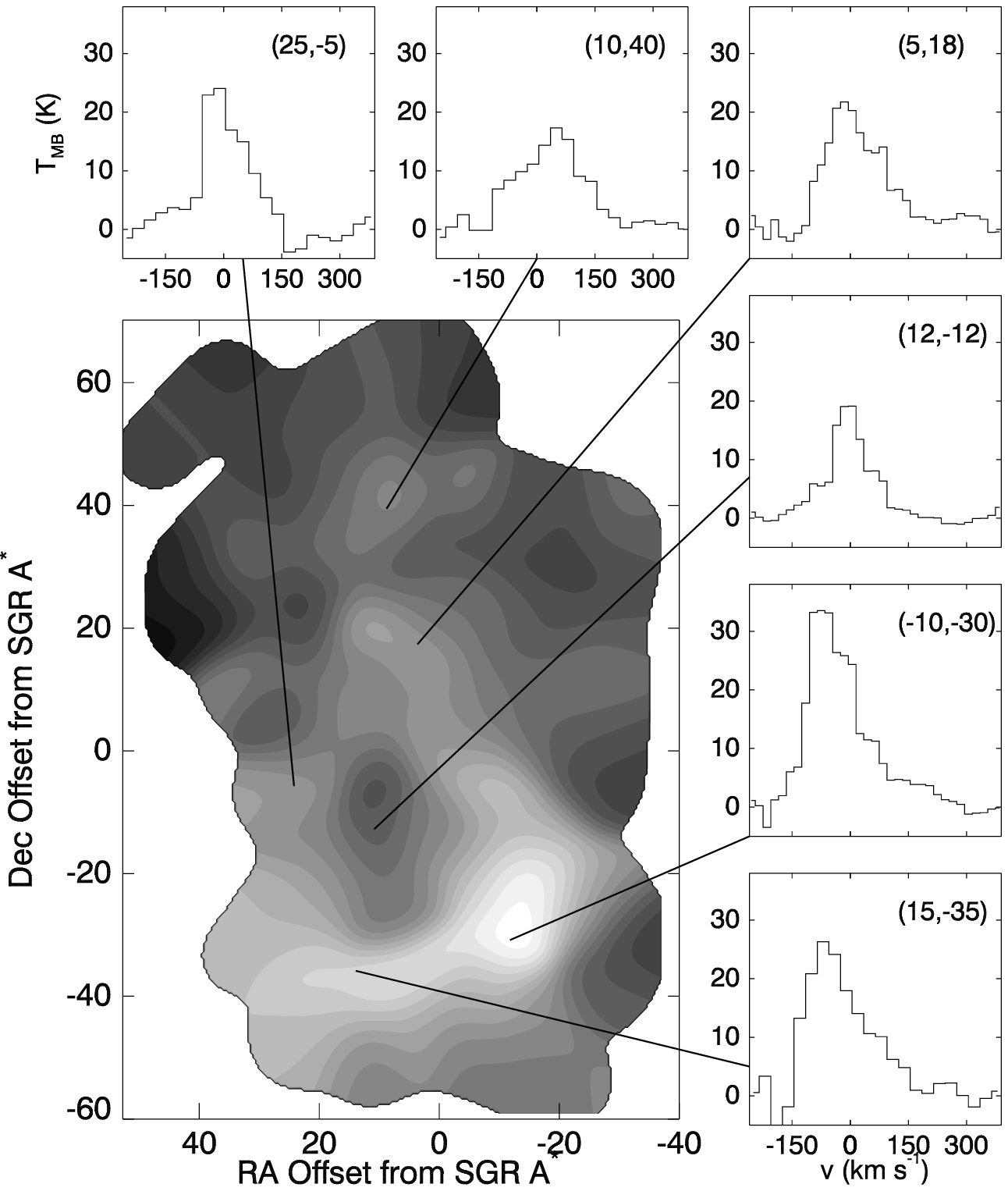}}}
\vskip1in
Figure 9, Bradford et al.  (single column)

\newpage
\centerline{\scalebox{0.9}{\includegraphics{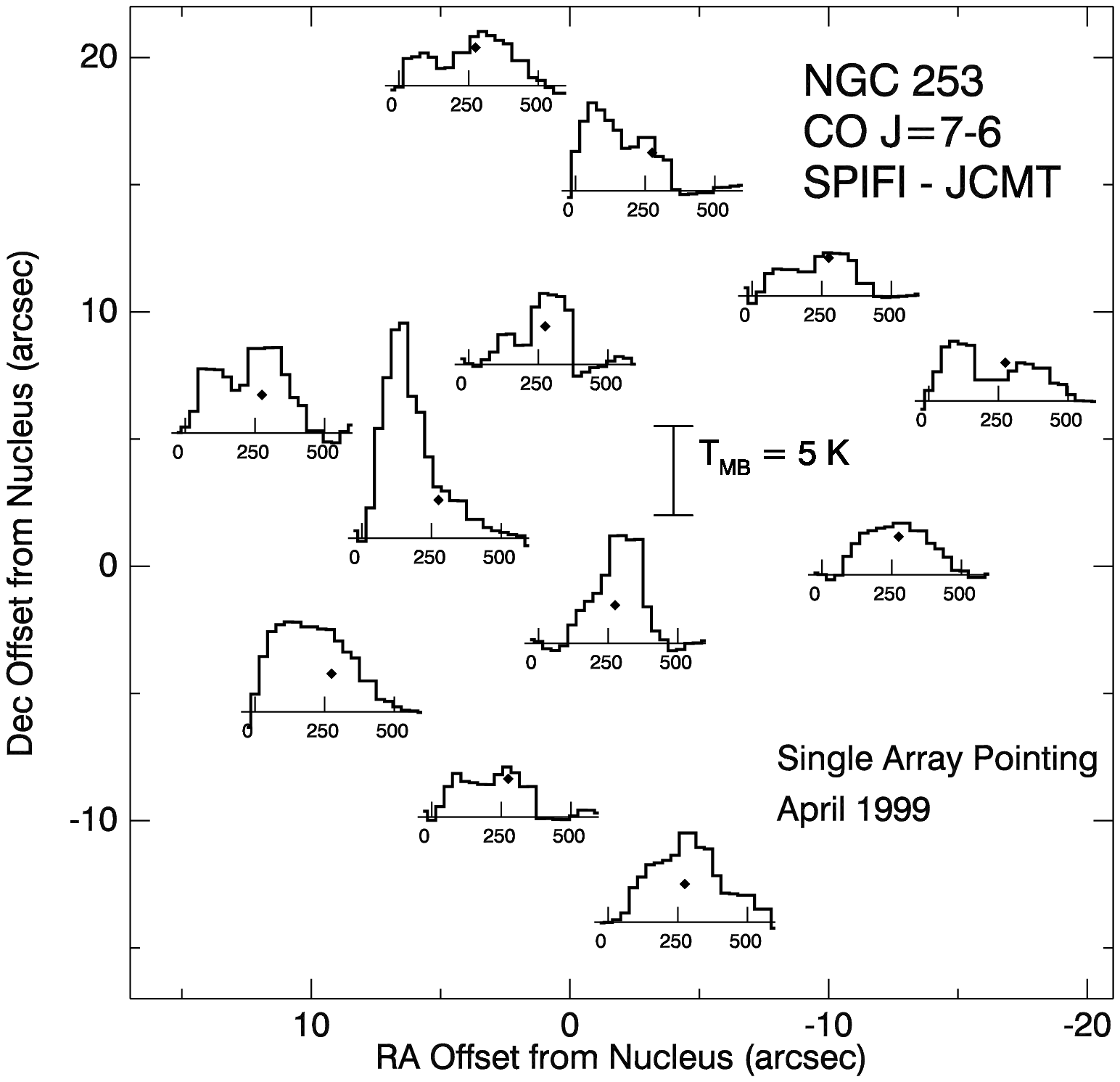}}}
\vskip1in
Figure 10, Bradford et al.  (single column)

\begin{table}[h]
\begin{center}
\caption{Elements in SPIFI Optical Train\label{tab:optics}} \vspace{0.1in}
\begin{tabular}{lclc}
\hline
Element & $T$ [K] & Spectral Property & Trans. at 370\mm \\
\hline \hline
PE Entrance Window & 280 & \nodata & 0.96 \\
Mirrors (six) & \nodata & \nodata & $(0.98)^6$ \\
High-Order FPI & 56 & $\frac{\lambda}{\Delta\lambda}\sim 500 - 10000$ & 0.70 \\
Lyot Stop Scatter Filter & 3.7 & $\lambda > 100$ \mm & 0.9 \\
Low-Order FPI & 3.7 & $\frac{\lambda}{\Delta\lambda}\sim 80 - 250$ & 0.75 \\
Low-Pass Mesh Filter & 1.5 & $\lambda > 200$ \mm & 0.85 \\
Bandpass Filter & 0.06 & $\frac{\lambda}{\Delta\lambda} = 8.8$ & 0.78 \\
Total & \nodata & $\frac{\lambda}{\Delta\lambda}\sim 500 - 10000$ & 0.26 \\

\hline
\end{tabular} \end{center} \end{table}

\clearpage

\begin{table}
\centering
\begin{minipage}{12.5cm}
\caption{SPIFI Detector Parameters}\label{tab:det}
\begin{tabular}{lcc}
\hline
\setcounter{footnote}{0} \renewcommand{\thefootnote}{\alph{footnote}}
\small \colhead{Parameter} & \colhead{Value} & \colhead{Unit}\\
\hline \hline
Bath Temperature & 60 & mK \\
Thermal Conductance  $G$& $2-3\e{-11}$ & $\rm W / K$\\
Operating Temperature & $80-110$ & mK \\
Dark Impedance & $5-100$ & $\rm M\Omega$ \\
Operating Impedance & $0.1-5$ & $\rm M\Omega$ \\
Optical Quantum Efficiency  $\eta$& 0.5 & \nodata \\
Electrical Responsivity $S_{e}$\footnotemark & $3-10 \times 10^8$ & ${\rm V / W}$ \\
Operating Johnson Noise\footnotemark & $3-10$ & ${\rm nV / \sqrt{Hz}}$ \\
Operating Thermal Noise & $3-7$ & ${\rm nV / \sqrt{Hz}}$  \\
Thermistor 1/f Noise at 8 Hz & $15-60$ & ${\rm nV / \sqrt{Hz}}$ \\
Amplifier Noise & 5 & ${\rm nV / \sqrt{Hz}}$ \\
Total non-photon contribution to NEV & $15-60$ & ${\rm nV / \sqrt{Hz}}$ \\
Total photon noise contribution to NEV & $35-40$ & ${\rm nV / \sqrt{Hz}}$ \\
\hline
\end{tabular}
\footnotetext[1]{Electrical responsivity does not include the optical quantum efficiency.}
\footnotetext[2]{All Noise Equivalent Voltages refer to the detector.}
\end{minipage}
\end{table}

\clearpage
\begin{table}[t]
\centering
\begin{minipage}{14cm}
\setcounter{footnote}{0} \renewcommand{\thefootnote}{\alph{footnote}}
\caption{SPIFI Sensitivities at JCMT\label{tab:sens}}\vspace{0.1in}
\begin{tabular}{ccccccc}
\hline
\small R $\left({\nu}/{\Delta\nu}\right)$&NEP\footnotemark&NEFD\footnotemark[1]&\tsys\footnotemark[2]
& ${\rm T_{sys}}$ PBGL \footnotemark[3] & Total BW\footnotemark[4]\\
  &$\rm\left[W/\sqrt{Hz}\right]$&$\rm\left[{Jy}/{\sqrt{Hz}}\right]$& [K] &
[K]&[\kms] \\ \hline \hline
10000 & $1.4 \times 10^{-14}$ & 196 & 3640 & 1450 & 400 \\ 
4800 & $1.7 \times 10^{-14}$ & 114 & 3060 & 1450 & 810 \\ 
3000 & $1.9 \times 10^{-14}$ & 80 & 2700 & 1450 & 1300 \\ 
1500 & $2.4 \times 10^{-14}$ & 50 & 2410 & 1450 &  2000 \\ 
800 & $3.1 \times 10^{-14}$ & 34 & 2280 & 1450 & 3100 \\ 

\hline
\end{tabular} \hspace{5cm}

\renewcommand{\baselinestretch}{1} \footnotesize Note: Sensitivities at the JCMT based on
measured values at R=4800 and 1500, scaled to good weather conditions.  Scanning is not
included; effective system temperatures including scanning are $\sqrt{n_{\rm res\;el}}$
larger, where $n_{\rm res\;el}$ is the number of spectral resolution elements in the scan.
\footnotetext[1]{The NEP refers to the main beam above the atmosphere (40\% transmission).
NEFD refers to a true point soure, also above the atmosphere.  Because these values
depend on the various beam efficiencies, they are strong functions of the antenna at the time of
observations.}  
\footnotetext[2]{System Temperatures referred to the forward beam above the atmosphere.}
\footnotetext[3]{System Temperature of a perfect background-limited (PBGL) spectrometer at
810 GHz on the JCMT.  This estimate assumes an instrument transmission of 25\%, detector
quantum efficiency of 50\%, re-imaging lenses with a transmission of 75\%, telescope forward
coupling of 65\%, and atmospheric transmission of 40\%}
\footnotetext[4]{Total bandwidth available in a single spectral setup.} 
\end{minipage} 

\end{table}


\begin{thebibliography}{}

\bibitem{koo00}J. Kooi, (2000), {\it personal communication}.

\bibitem{hil00}R. Hills, J. Richer, S. Withington, H. Smith, H. Gibson, B. Dent, W. Duncan, J. Harris, P. Hastings,  L. Avery, C. Cunningham, P. Feldman, R. Redman, K. Yeung and P. Jewell, ``Heterodyne array receiver programme for the James Clerk Maxwell Telescope,'' \url{http://www.jach.hawaii.edu/JACpublic/JCMT/harp-info.html}, (2000).

\bibitem{kee85}J. Keene, G.A. Blake, T.G. Phillips, P.J. Huggins and C.A. Beichman, ``The abundance of atomic carbon near the ionization fronts in M17 and S140,'' \apj {\bf 299}, 967--980 (1985).

\bibitem{zmu88}J. Zmuidzinas, A.L. Betz, R.T. Boreiko and D.M. Goldhaber, ``Neutral atomic carbon in dense molecular clouds,'' \apj {\bf 335}, 774--785 (1988).

\bibitem{wal93}C.K. Walker, G. Narayanan, T. Buttgenbach, J. Carlstrom, J. Keene and T.G. Phillips, ``The detection of [CI] in molecular outflows associated with young stellar objects,'' \apj, {\bf 415}, 672--679 (1993).

\bibitem{har99}Harris, A.I., ``Directions for Submillimeter and Far-Infrared
Instrumentation,'' {\it The Physics and Chemistry of the Interstellar Medium, Proceedings
of the 3rd Cologne-Zermatt Symposium}, V. Ossenkopf, J. Stutzki, and G. Winnewisser, eds.
(GCA-Verlag Herdecke, 1999)

\bibitem{lam86}J.M. Lamarre,  ``Photon noise in photometric instruments at far-infrared
and submillimeter wavelengths,'' \ao {\bf 6}, 870-876 (1986)

\bibitem{plu94}R. Plume, D.T. Jaffe, and J. Keene, ``Observations of large-scale [CI] emission from S140,'' \apj, {\bf 425}, L49--L52 (1994).

\bibitem{wil92}W. Wild, A.I. Harris, A. Eckhart, U.U. Graf, J.M. Jackson, A.P.G. Russell and J. Stutzki, ``A multi-line study of the molecular interstellar medium in M 82's starburst nucleus,'' \aaa, {\bf 265}, 447--464 (1992).

\bibitem{cz96}J.E. Carlstrom and J. Zmuidzinas, ``Millimeter and submillimeter techniques,'' in {\it Reviews Radio Science 1993-1995}, W.R. Stone, ed.  (Oxford University Press, Oxford , 1996).

\bibitem{gp98}M. Gerin and T.G. Phillips, ``Atomic carbon in Arp 220,'' \apj {\bf 509}, L17--L20 (1998).

\bibitem{sta91}G.J. Stacey, J.W. Beeman, E.E. Haller, N. Geis, A. Poglistch and M. Rumitz, ``Stressed and unstressed Ge:Ga photoconductor arrays for far-IR astronomy,'' Int. J. of IR and MM Waves, {\bf 13}, 1689--1700 (1991)

\bibitem{hol98}W.S. Holland, C.R. Cunningham, W.K. Gear, T. Jenness, K. Laidlaw, J.F. Lightfoot and E.I. Robson, ``SCUBA, a submillimeter camera operating on the James Clerk Maxwell Telescope,'' in {\it Advanced Technology MMW, Radio, and Terahertz Telescopes}, T.G. Phillips, ed., Proc. SPIE {\bf 3357}, 305--318 (1998).

\bibitem{wan96}N. Wang, T.R. Hunter, D.J. Benford, E. Serabyn, D.C. Lis, T.G. Phillips,
S.H. Moseley, K. Boyce, A. Szymkowiak, C. Allen, B. Mott, and J. Gygax, ``Characterization of a submillimeter high-angular-resolution camera with a monolithic silicon bolometer array for the Caltech Submillimeter Observatory,'' \ao {\bf 35}, 6629--6640 (1996).

\bibitem{swa98}M.R. Swain, C.M. Bradford, G.J. Stacey, A.D. Bolatto, J.M. Jackson, M. Savage, J.A. Davidson, ``Design of the South Pole Imaging Fabry-Perot Interferometer (SPIFI),'' in {\it Infrared Astronomical Instrumentation}, A. Fowler, ed., Proc. SPIE {\bf 3354}, 480--492 (1998).

\bibitem{lat97}H.M. Latvakoski, ``High spatial resolution mid and far-infrared imaging of the Galactic Center,'' PhD Thesis, Cornell University (1997).

\bibitem{tf85}John R. Tucker and Marc J. Feldman, ``Quantum detection at millimeter
wavelengths,''  Reviews of Modern Physics, {\bf 57}, 1055--1113 (1985).

\bibitem{pog91}A. Poglitsch, J.W. Beeman, N. Geis, M. Haggerty, E.E. Haller, J.M. Jackson,
M. Rumitz, G.J. Stacey, and C.H. Townes, ``The MPE / UCB far-infrared imaging Fabry-Perot interferometer (FIFI),'' Int. J. of IR and MM Waves, {\bf 12}, 859--870 (1991).

\bibitem{har76}A. Harper, R.H. Hildebrand, R. Stiening and R. Winston, ``Heat trap -- An optimized far infrared field optics system,'' \ao {\bf 15}, 53--60 (1976).

\bibitem{sch87}D.J. Schroeder, {\it Astronomical Optics} (Academic Press, San Diego, 1987), Chap. 15.

\bibitem{bw80}M. Born and E. Wolf, {\it Principles of Optics} (Pergamon Press, Oxford, 1980), Chap. 7.

\bibitem{sta93}G.J. Stacey, T.L. Hayward, H.M. Latvakoski and G.E. Gull, ``KWIC: A widefield mid-infrared array camera / spectrometer for the KAO,'' in {\it Infrared Detectors and Instrumentation}, A.M. Fowler, ed., Proc. SPIE {\bf 1946}, 238--248 (1993).

\bibitem{ash91}J. Ashok, P.L.H. Varaprasad, and J.R. Birch, ``Polyethylene,'' in {\it Handbook of Optical Constants of Solids II}, Edward D. Palik, ed. (Academic Press, Boston, 1991).

\bibitem{sg83}K. Sakai and L. Genzel, ``Far-infrared metal mesh filter and Fabry-Perot interferometry,'' in {\it Reviews of Infrared and Millimeter Waves}, K.J. Button, ed. (Plenum, New York, 1983).

\bibitem{mcc93}D. McCammon, W. Cui, M. Juda, J. Morgenthaler, J. Zhang, R.L. Kelley, S.S. Holt, G.M. Madejski, S.H. Moseley and A.E. Szymkowiak, ``Thermal calorimeters for high resolution X-ray spectroscopy,'' Nucl. Inst. and Meth. in Phys. Res., {\bf A326}, 157--165 (1993).

\bibitem{mos84}S.H. Moseley, J.C. Mather and D. McCammon, ``Thermal detectors as X-ray spectrometers,'' J. Appl. Phys. {\bf 56}, 1257--1262 (1984).

\bibitem{mat82}J.C. Mather, ``Bolometers:  ultimate sensitivity, optimization, and amplifier coupling,'' \ao {\bf 23}, 584--588 (1982).

\bibitem{han98}S.-I. Han, R. Almy, E. Apodaca, W. Bergmann, S. Deiker, A. Lesser, D. McCammon, K. Rawlins, R. L. Kelly, S.H. Moseley, F.S. Porter, C.K. Stahle and A.E. Szymkowiak, ``Intrinsic 1/f noise in doped silicon thermistors for cryogenic calorimeters,'' in {\it EUV, X-Ray and Gamma Ray Instrumentation for Astronomy IX}, O.H. Seigmund and M.A. Gummin, eds., Proc. SPIE {\bf 4435}, 640--644 (1998).

\bibitem{whi79}G.K. White, {\it Experimental Techniques in Low-Temperature Physics} (Clarendon Press, Oxford, 1979), Chap. 9.

\bibitem{fis98}H.E. Fischer, ``Magnetic cooling,'' in {\it Experimental Techniques in Condensed Matter Physics at Low Temperatures}, R.C. Richardson and E.N. Smith, eds.  (Addison-Wesley, Reading, MA, 1998).

\bibitem{mcc96}D. McCammon, R. Almy, S. Deiker, J. Morgenthaler, R.L. Kelley, F.J. Marshall, S.H. Moseley, C.K. Stahle and A.E. Szymkowiak, ``A sounding rocket payload for X-ray astronomy employing high-resolution microcalorimeters,'' Nucl. Inst. and Meth. in Phys. Res., {\bf A 370}, 266--268 (1996).

\bibitem{par01a}J.R. Pardo, J. Cernicharo, and E. Serabyn, ``Atmospheric transmission at
microwaves (ATM): an improved model for mm/submm applications,''  IEEE Trans. on Ant. and
Prop., in press.

\bibitem{par01b}J.R. Pardo, J. Cernicharo, and E. Serabyn, ``Submillimeter atmospheric
transmission measurements on Mauna Kea during extremely dry El Nino conditions:
implications for broadband opacity contributions,'' Journal of Quantitative Spectroscopy
and Radiative Transfer, {\bf 68:}(4) 419-433 (2001).

\bibitem{sta01a}R. Stark, {\it Personal Communication}, 

\bibitem{sta01b}R. Stark, ``MPIRE--the MPIfR/SRON 800 GHz heterodyne spectrometer,''
\url{http://www.mpifr-boonn.mpg.de/div/mm/tech/mpire.html}, 2000.

\bibitem{mat99}H. Matthews, ``Estimating time requirements and sensitivity for heterodyne receivers,'' \url{www.jach.hawaii.edu/JACpublic/JCMT/userguide/guide}, (1999).

\bibitem{zmu00}J. Zmuidzinas, {\it manuscript in preparation} (2000).

\bibitem{ben98}D.J. Benford, T.R. Hunter, \& T.G. Phillips, ``Noise equivalent power of background limited thermal detectors at submillimeter wavelengths,''  Int. J. of IR \& MM Waves, {\bf 19}, 931--938 (1998).

\bibitem{koo00b}J.W. Kooi, J. Kawamura, J. Chen, G. Chattopadhyay, J.R. Pardo, J. Zmuidzinas, T.G. Phillips, B. Bumble, J. Stern \& H.G. LeDuc, ``A Low-Noise, NbTiN-based 850 GHz SIS Receiver for the Caltech Submillimeter Observatory,'' Int. J. of IR \& MM Waves, {\bf 21}, XX  (2000).

\bibitem{rie94}Rieke, G.H., {\it Detection of Light from the Ultraviolet to the
Submillimeter} (Cambridge University Press, Cambridge, MA, 1994), Chap. 9.

\end{thebibliography}
\end{document}